\documentclass[a4paper,11pt]{article}
\pdfoutput=1 
\usepackage{jcappub}
\usepackage[utf8]{inputenc}
\usepackage[normalem]{ulem}
\usepackage{subfigure}
\usepackage{epsfig} 
\usepackage{amsmath} 
\usepackage{amsfonts}  
\usepackage{float}    
\usepackage{amssymb}   
\usepackage{slashed} 
\usepackage{placeins}
\usepackage{mathrsfs}
\usepackage{color}       
\usepackage{pbox}
\usepackage{array,multirow,makecell}
\usepackage{natbib} 
\usepackage{lipsum}
\usepackage{appendix}

\newcommand{\lsim}[1]{\lesssim}

\makeatletter
\gdef\@fpheader{}
\makeatother

\newcommand{\diff}{\mathop{}\!\mathrm{d}}

\usepackage{orcidlink}
\definecolor{DodgerBlue}{RGB}{30,144,255}
\begin{document}
\title{Primordial black holes as cosmic accelerators of light dark matter: Novel direct detection constraints}

\author[a,b]{Sk Jeesun~\orcidlink{0009-0005-2344-9286},}
\emailAdd{jeesun@sjtu.edu.cn}
\affiliation[a]{State Key Laboratory of Dark Matter Physics, Tsung-Dao Lee Institute \& School of Physics and Astronomy, Shanghai Jiao Tong University, Shanghai 200240, China}
\affiliation[b]{Key Laboratory for Particle Astrophysics and Cosmology (MOE) \& Shanghai Key Laboratory for Particle Physics and Cosmology, Shanghai Jiao Tong University, Shanghai 200240, China}

\author[c]{Anirban Majumdar~\orcidlink{0000-0002-1229-7951},}
\emailAdd{anirban19@iiserb.ac.in}
\affiliation[c]{Department of Physics, Indian Institute of Science Education and Research - Bhopal, \\ 
Bhopal Bypass Road, Bhauri, Bhopal 462066, India}

\author[c]{Rahul Srivastava~\orcidlink{0000-0001-7023-5727}}
\emailAdd{rahul@iiserb.ac.in}

\abstract{Current multi-tonne-scale dark matter (DM) detectors are largely incapable of detecting light dark matter from the Galactic halo due to the energy threshold limitations of their recoil measurements. However, primordial black holes (PBHs) can evaporate via Hawking radiation to particles whose energies are set by the black hole temperature. Consequently, weakly interacting light dark matter (or dark radiation) particles produced in this manner can reach the Earth with sufficient flux and kinetic energy above the experimental thresholds. This  opens up a novel avenue to probe the light dark sector in terrestrial experiments. In this work, we explore this possibility by considering fermionic DM produced through PBH evaporation and investigating its electron recoil signatures in direct detection experiments. 
We analyze both energy independent (constant) and energy dependent (scalar and vector mediated) DM-electron interactions, highlighting the strong dependence of the recoil spectra on the underlying Lorentz structure of the interaction. In addition, we also account for the attenuation effects due to the loss of kinetic energy while DM traverses through Earth's crust, which can significantly modify the incoming DM flux. Incorporating these effects carefully, we place constraints on light DM using the electron recoil data from XENONnT, LZ, and PandaX-4T. Finally, we also discuss the detection prospects of such dark matter in current and future generation neutrino detectors, such as Super-Kamiokande and Hyper-Kamiokande.
}

\maketitle

\section{Introduction}

Numerous astrophysical and cosmological observations indicate that non-luminous and non-baryonic dark matter (DM) constitutes about $27\%$ of the total energy density of the Universe, corresponding to nearly $84\%$ of its matter content~\cite{Zwicky:1933gu,Rubin:1970zza,Clowe:2006eq,Planck:2018vyg}. Despite overwhelming gravitational evidence, the microscopic nature and fundamental particle properties of DM remain unknown. A widely studied and long-standing paradigm is that of thermal weakly interacting massive particles (WIMPs), whose relic abundance is set by weak-scale interactions. However, no conclusive signal of WIMP DM has been identified to date in direct detection experiments, despite decades of increasingly sensitive searches~\cite{Bertone:2004pz, Queiroz:2016awc}. This situation has motivated growing interest in scenarios that depart from the conventional WIMP framework, leading to a broad exploration of alternative possibilities. These include models with light (sub-GeV) particle DM~\cite{Griest:1990kh, Essig:2011nj, Essig:2017kqs, Elor:2021swj, Gninenko:2023pkv} as well as non-thermal, macroscopic compact objects such as primordial black holes (PBHs)~\cite{Barrau:2003xp, Carr:2020gox, Villanueva-Domingo:2021spv}.

Among the various alternatives to particle DM, PBHs have emerged as a particularly intriguing possibility~\cite{Belotsky:2014kca, Green:2020jor, GilChoi:2023ahp}. Unlike astrophysical black holes, which originate from the collapse of massive stars, PBHs could have formed in the early Universe through a different mechanism, such as the gravitational collapse of large density inhomogeneities~\cite{Carr:1993aq, Ivanov:1994pa, Garcia-Bellido:1996mdl, Randall:1995dj, Khlopov:2008qy}. As a result, their masses are not restricted to stellar scales and can span many orders of magnitude, ranging from well below to far above solar masses~\cite{Carr:1975qj}. The idea dates back to early work by Zel’dovich and Novikov~\cite{Zeldovich:1967lct}, and was later developed extensively by Hawking and Carr~\cite{10.1093/mnras/152.1.75, Carr:1974nx}, stimulating an exhaustive theoretical and observational investigation in this direction. PBHs can in principle constitute a fraction, or even the entirety, of the DM, and numerous observational efforts have been directed towards constraining this possibility~\cite{Casanueva-Villarreal:2025kmd}. A particularly interesting aspect of PBHs is that, unlike conventional DM candidates, they are expected to emit radiation due to quantum effects near their event horizon. This process, known as Hawking evaporation, causes PBHs to gradually lose mass and produce a spectrum of particles~\cite{Hawking:1974rv, Hawking:1975vcx}. For sufficiently light PBHs, the evaporation is efficient on cosmological timescales. In particular, PBHs in the approximate mass range $7\times10^{14}\!-\!10^{19}\,\mathrm{gm}$ are subject to strong constraints from their evaporation products, while PBHs lighter than $\sim 7\times10^{14}\,\mathrm{gm}$ would have completely evaporated by today, leaving only a redshifted background of the particles they emitted~\cite{Arbey:2021mbl}. These evaporation products therefore provide a powerful indirect probe of the possible existence of PBHs. Among the resulting emission channels, photons and charged leptons are subject to a wide range of cosmological and astrophysical constraints~\cite{Carr:2020gox, Barrau:2003xp, Laha:2019ssq, Laha:2020ivk, Saha:2021pqf, Ray:2021mxu, Capanema:2021hnm, Wu:2025ovd}, while neutrino signals from evaporating PBHs have also been extensively explored~\cite{Dasgupta:2019cae,Wang:2020uvi,Calabrese:2021zfq,DeRomeri:2021xgy,Capanema:2021hnm,Bernal:2022swt, Wu:2024uxa}. In addition to these Standard Model (SM) species, PBHs may serve as an efficient source of relativistic or semi-relativistic light DM particles~\cite{Morrison:2018xla,Baldes:2020nuv,Bernal:2020kse,Bernal:2020ili,Bernal:2020bjf,Auffinger:2020afu,Gondolo:2020uqv,Masina:2021zpu,Cheek:2021odj,Cheek:2021cfe}. For PBHs with masses around $10^{15}\,\mathrm{gm}$-$10^{17}\,\mathrm{gm}$, the emitted DM particles can reach kinetic energies up to hundreds of MeV, allowing them to traverse the Earth and potentially be detected in terrestrial experiments.

The current generation of multi-tonne-scale direct detection (DD) experiments has reached remarkable sensitivity to electroweak-scale DM, particularly for masses $\gtrsim 5~\mathrm{GeV}$. However, at lower masses, the expected nuclear and electron recoil energies become too small to exceed detector thresholds, leading to a rapid loss of sensitivity. In the standard halo scenario, non-relativistic MeV/sub-MeV DM with typical Galactic velocities $v\sim 10^{-3}$ is very difficult to detect. By contrast, if a small fraction of the DM population is accelerated to (semi-) relativistic energies, it can transfer sufficient momentum to produce observable recoil signals. Such energetic DM would constitute only a subdominant component of the total DM density, yet it can significantly extend the reach of DD experiments into the MeV/sub-MeV regime. This idea was first explored in scenarios where DM is boosted through scattering with Galactic cosmic rays~\cite{Bringmann:2018cvk,Yin:2018yjn,Ema:2018bih}, and has since motivated a wide range of theoretical developments and phenomenological studies~\cite{Cappiello:2018hsu,Dent:2019krz,Alvey:2019zaa,Lei:2020mii,Dent:2020syp,Xia:2021vbz,CDEX:2022fig,Alvey:2022pad,Arguelles:2022fqq,Maity:2022exk,Cappiello:2019qsw,Jho:2020sku,Cho:2020mnc,PandaX-II:2021kai,Su:2023zgr,Herrera:2023nww, Ghosh:2024dqw, Herbermann:2024kcy, Jeesun:2026lro}. More recently, boosted DM scenarios arising from interactions with neutrinos have been proposed, including cosmic, solar, and supernova neutrinos, as well as neutrinos from PBH evaporation or the diffuse supernova neutrino background~\cite{Jho:2021rmn,Zhang:2020nis,Chao:2021orr,Calabrese:2021zfq,Calabrese:2022rfa,Li:2022jxo,Lin:2022dbl, DeRomeri:2023ytt,Lin:2023nsm,Das:2021lcr,Ghosh:2021vkt, Das:2024ghw}. 
Additional mechanisms have also been discussed, such as DM boosted by blazars or by cosmic rays confined in starburst galaxies, as well as boosted DM originating from vacuum decay in phantom dark-energy scenarios, and scenarios where interactions between light DM and relativistic cosmic rays in astrophysical jets modify the multi-wavelength emission from Active Galactic Nuclei (AGN), or arising in semi-annihilating and multi-component DM frameworks~\cite{Bhowmick:2022zkj,Cline:2023hfw, Kantzas:2025huu,Jeesun:2025gzt,Ambrosone:2022mvk,DEramo:2010keq,Berger:2014sqa,Toma:2021vlw,Agashe:2014yua,Kong:2014mia,Aoki:2018gjf,Borah:2021yek}.
As discussed in the last paragraph, PBHs themselves naturally provide an efficient mechanism for producing (semi-) relativistic light DM through Hawking evaporation, and this possibility has recently begun to be investigated in the context of DD experiments~\cite{Calabrese:2022rfa, Calabrese:2021src, Li:2022jxo, CDEX:2024xqm, Arcadi:2024tib, DarkSide-50:2025olf}.

In this work, we investigate the prospects for detecting light DM produced via PBH evaporation through DM--electron scattering in terrestrial experiments. We make use of the electron recoil data sets from XENONnT~\cite{XENON:2022ltv}, LZ~\cite{LZ:2022lsv}, and PandaX-4T (Run~1)~\cite{PandaX:2024cic} to constrain the DM--electron scattering cross-section and the DM mass. These experiments employ state-of-the-art liquid xenon detectors designed to observe very small electronic and nuclear recoil signatures. XENONnT located at the underground Gran Sasso National Laboratory in Italy, represents the upgraded phase of XENON1T and has reported data with a total exposure of $1.16~\mathrm{tonne}\times\mathrm{year}$. The LZ experiment, operating at the Sanford Underground Research Facility in South Dakota, has provided data corresponding to an exposure of $5.5~\mathrm{tonne}\times60~\mathrm{days}$, while the PandaX-4T detector at the China Jinping Underground Laboratory has recently released its Run~1 results with an exposure of $363.3~\mathrm{tonne\cdot day}$. Thanks to their large target masses and high sensitivity, excellent photon collection capabilities, and very low backgrounds, these experiments offer highly complementary and competitive searches for weakly interacting DM.

Previous studies of PBH-evaporated DM typically assumed a constant DM--electron scattering cross-section~\cite{Li:2022jxo}, which is only valid for non-relativistic scattering through heavy mediators. Since the DM emitted from PBHs is (semi-) relativistic, such an approximation may not always be justified. In addition to the conventional constant cross-section case, in this work we therefore incorporate the full relativistic DM-electron scattering cross-section for scalar and vector mediators leading to spin-independent interactions, representing the first application of this treatment to PBH-evaporated DM. As we show in later sections, the limits obtained for scalar and vector mediators differ significantly from that obtained assuming just a constant cross-section.

Moreover, DM particles produced via PBH evaporation can lose kinetic energy while propagating through the atmosphere and Earth before reaching underground detectors. Ref.~\cite{Calabrese:2022rfa} indeed explores an energy-dependent cross-section but considers only scalar mediated cross-section and doesn't consider the attenuation effect. 
On the other hand,  Ref.~\cite{Li:2022jxo} relied on simplified analytic estimates to consider the attenuation effect whose validity breaks down when the DM kinetic energy is comparable to the electron mass.
In that direction, we improve upon earlier analyses performed in Ref.~\cite{Li:2022jxo,Calabrese:2022rfa} by implementing a fully numerical calculation of this in-medium attenuation. By solving for the mean energy loss with the approximation of negligible deflection, we quantify the resulting modifications to the arriving DM flux at the underground detectors and to the derived constraints using the updated data from DD experiments.
Alternatively, we also derive limits on the PBH abundance parameter $f_{\rm PBH}$ as a function of PBH mass for a few benchmark DM masses and cross-sections, as an example of how PBH evaporated dark particle can help to probe PBH complementary to the
bounds obtained from extragalactic gamma-ray measurements~\cite{Carr:2020gox}.

Finally, for completeness, we also explore the possibility of probing such PBH evaporated light particles in the current and future generation multi-tonne-scale neutrino detectors. As an example we consider Super-Kamiokande \cite{Super-Kamiokande:2017dch} and the future generation experiment Hyper-Kamiokande \cite{Hyper-Kamiokande:2018ofw}.
Super-K is a water Cherenkov detector with 22.5 kilotonne fiducial mass and its upgrade Hyper-K with a fiducial mass of 187 kilotonne is expected to start in near future as a next generation neutrino detector. 
Despite their high thresholds, these detectors benefit from their larger target volume compared to DM detectors and hence, they have been able to place stringent constraints on light boosted DM \cite{Ema:2018bih,Super-Kamiokande:2017dch}.
Using Super-K and Hyper-K we obtain conservative limits on light DM evaporated from PBH assuming energy dependent cross-sections.
At this stage it is worth highlighting that, though we refer the emitted particle as DM it can be a generic dark sector particle (dark radiation) as well.
Thus in this work we present a  detailed analysis of some of the terrestrial experiments, such as multi-tonne-scale DM direct detection experiments  (XENONnT, LZ, and PandaX-4T), and large-scale neutrino detectors (Super-K and Hyper-K), which can probe such PBH evaporated dark sector particles and  provide competitive, and in certain regimes stronger, constraints compared to existing sub-GeV DM bounds.

The remainder of this paper is organized as follows. Section~\ref{Sec:PBH_Flux} provides a brief discussion on the estimation of PBH evaporated DM flux. In Sec.~\ref{Sec:Event_Simulation}, we detail the simulation of the PBH-evaporated DM signal at the XENONnT, LZ, and PandaX-4T detectors, including the calculation of recoil spectra, detector response, and the statistical framework used to analyse the direct detection data. In Sec.~\ref{sec:result_unatt}, we showcase the resulting constraints without invoking DM attenuation. Sec.~\ref{Sec:Attenuation_Technique} contains a detail discussion of Earth's attenuation effects on the incoming boosted DM particles. 
In Sec.~\ref{sec:results}, we display the possible recoil spectra and constraints on PBH evaporated DM from XENONnT, LZ and PandaX-4T after considering attenuation effect. 
In Sec.~\ref{sec:fpbh} we showcase the possible indirect limit from DM detectors on PBH parameters through the (null) observation of any dark sector particle evaporated from it.
The existing limit and projected reach of neutrino experiments are also discussed in Sec.~\ref{sec:nu_detector} and finally we conclude in Sec.~\ref{sec:concl}.

\section{Estimation of dark matter flux evaporated from PBH}\label{Sec:PBH_Flux}
Primordial black holes (PBHs), formed in the early Universe from the collapse of large density fluctuations, can emit particles through Hawking radiation arising from quantum effects near their event horizon. The resulting evaporation is a purely gravitational process that produces a nearly thermal spectrum of all kinematically allowed particle species, characterized by the Hawking temperature~\cite{Hawking:1975vcx},  
\begin{equation}
T_\mathrm{PBH} = \frac{\hbar c^{3}}{8\pi G M_\mathrm{PBH} k_\mathrm{B}} 
\approx 10^{-7} \left( \frac{M_\mathrm{PBH}}{M_\odot} \right)^{-1} \, \text{K} ,
\end{equation}
where \( G \) is the Newtonian constant of gravitation, \( \hbar \) is the reduced Planck constant, \( c \) is the speed of light, \( k_\mathrm{B} \) is the Boltzmann constant, \( M_\mathrm{PBH} \) denotes the PBH mass, and \( M_\odot \) is the solar mass. This formula signifies that the Hawking temperature is inversely proportional to the PBH mass. Hence, at early times, when the temperature of the PBH is relatively low, only light particles such as photons and neutrinos are efficiently emitted. As the PBH mass decreases with time, its temperature rises, leading to the evaporation of heavier species, potentially including particles from the dark sector.

The differential emission rate of dark matter particles from a PBH is described by~\cite{Hawking:1975vcx}
\begin{equation}
    \label{Eq:DM_number_Spectra}
    \frac{\diff^{2} N_\chi}{\diff T_\chi\, \diff t}
    = \frac{g_\chi}{2\pi}\,
      \frac{\Gamma(T_\chi,\,M_\mathrm{PBH})}
           {\exp\!\left[(T_\chi + m_\chi)/(k_\mathrm{B} T_\mathrm{PBH})\right] + 1}\,,
\end{equation}
where $g_\chi$ and $T_\chi$ denote the degrees of freedom and kinetic energy of the emitted DM particle, respectively, while $m_\chi$ denotes its mass. In the present analysis we assume the emitted DM to be a Dirac fermion, hence $g_\chi=4$, although the formalism can be straightforwardly extended to scalar, vector, or Majorana DM candidates. The function $\Gamma(T_\chi,\,M_\mathrm{PBH})$ is the graybody factor, which quantifies the probability that an emitted spin-$1/2$ DM particle surpasses the PBH gravitational potential barrier and escapes to infinity\footnote{In general the graybody factor also depends on additional PBH parameters, such as spin ($a_\mathrm{PBH}$) or charge ($Q_\mathrm{PBH}$); however, in this work we restrict to static, uncharged Schwarzschild PBHs with $a_\mathrm{PBH}=0$ and $Q_\mathrm{PBH}=0$.}. In our numerical calculations, $\Gamma$ is obtained using the publicly available \texttt{BlackHawk v2.3} code~\cite{Arbey:2021mbl}.

For the DM flux originating from PBH evaporation, we include contributions from both PBHs residing in the Milky Way halo and those distributed throughout the extragalactic Universe~\cite{Wang:2020uvi},
\begin{eqnarray}
\frac{\diff^2\Phi_\chi}{\diff T_\chi\, \diff\Omega}
= \frac{\diff^2\Phi_\chi^{\rm MW}}{\diff T_\chi\, \diff\Omega}
+ \frac{\diff^2\Phi_\chi^{\rm EG}}{\diff T_\chi\, \diff\Omega}\,,
\label{eq:PBH_Flux}
\end{eqnarray}
where $\Phi_\chi^{\rm MW}$ and $\Phi_\chi^{\rm EG}$ denote the fluxes associated with Galactic and extragalactic PBHs, respectively, and $\Omega$ is the solid angle.
\begin{figure}[tbh!]
    \centering
    \includegraphics[scale=0.5]{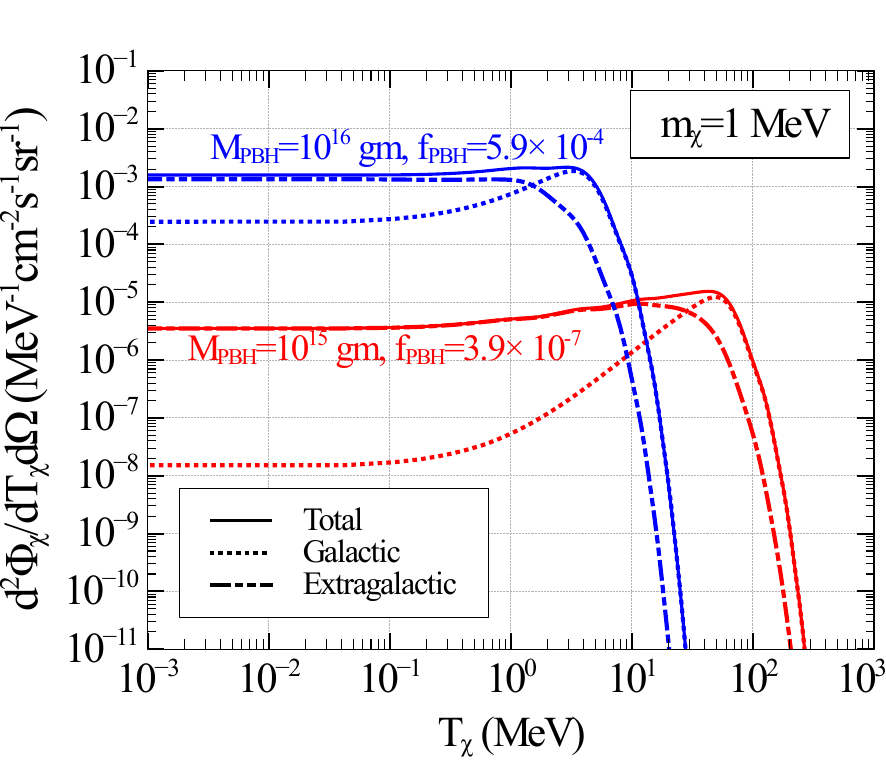}
    \caption{Differential DM flux per unit solid angle, $\diff^{2}\Phi_\chi/\diff T_\chi\,\diff\Omega$, from PBH evaporation as a function of the DM kinetic energy $T_\chi$ for a DM mass of $m_\chi = 1~\mathrm{MeV}$. Results are shown for two benchmark PBH masses, $M_{\rm PBH}=10^{15}~\mathrm{gm}$ (red) and $10^{16}~\mathrm{gm}$ (blue), with corresponding PBH abundance fractions consistent with cosmological limits, $f_{\rm PBH}=3.9\times10^{-7}$ and $5.9\times10^{-4}$, respectively. For each benchmark, the Galactic (dashed), extragalactic (dash–dotted), and total (solid) flux components are displayed.}
    \label{fig:uan_flux}
\end{figure}
The Galactic contribution is given by
\begin{eqnarray}
\frac{\diff^2\Phi_\chi^{\rm MW}}{\diff T_\chi\, \diff\Omega}
= D_{\rm halo}\,\frac{f_{\rm PBH}}{4\pi\,M_{\rm PBH}}
\left.\frac{\diff^2 N_\chi}{\diff T_\chi\,\diff t}\right|_{t=t_{\rm univ}}\,,
\label{eq:MW_DM_flux}
\end{eqnarray}
where $t_{\rm univ}\approx 13.6~\mathrm{Gyr}$ is the age of the Universe and 
$f_{\rm PBH}\equiv\Omega_{\rm PBH}/\Omega_{\rm DM}$ is the fraction of DM in the form of PBHs.  
The quantity $D_{\rm halo}$ encapsulates the line-of-sight (l.o.s.) integral of the Galactic DM density profile over the region of interest,
\begin{equation}
\label{eq:Dfactor}
D_{\rm halo}
= \int_{\Delta\Omega} \frac{\diff\Omega}{4\pi}
\int_0^{\ell_{\rm max}} \rho_{\rm MW}[r(\ell,\psi)]\,\diff\ell\,.
\end{equation}
Throughout this work we adopt a Navarro--Frenk--White (NFW) profile~\cite{Navarro:1996gj},
\begin{equation}
\label{equn:NFW}
\rho_{\rm MW}(r)
= \rho_\odot
\left(\frac{r}{r_\odot}\right)^{-1}
\left[\frac{1 + \frac{r_\odot}{r_s}}{1 + \frac{r}{r_s}}\right]^2 ,
\end{equation}
with scale radius $r_s = 20~\mathrm{kpc}$ and local DM density $\rho_\odot = 0.4~\mathrm{GeV\,cm}^{-3}$.  
$r$ is the galactocentric distance along the l.o.s. which reads as,
\begin{equation}
r(\ell,\psi)
= \sqrt{r_\odot^2 - 2\ell r_\odot \cos\psi + \ell^2}\,,
\end{equation}
where $r_\odot = 8.5~\mathrm{kpc}$ is the Sun’s distance from the Galactic centre and $\psi$ is the observation angle.  
The upper limit of the integration in eq.~\eqref{eq:Dfactor} is given by,
\begin{equation}
\ell_{\rm max}
= \sqrt{R^2 - r_\odot^2\sin^2\psi} + r_\odot\cos\psi\,,
\end{equation}
with the halo virial radius taken as $R=200~\mathrm{kpc}$.  
Using these parameters, we obtain 
\(
D_{\rm halo}=2.22\times 10^{25}~\mathrm{MeV\,cm^{-2}}
\)
for the full Milky Way halo.
On the other hand, the extragalactic flux contribution over the full sky reads
\begin{eqnarray}
\frac{\diff^2\Phi_\chi^{\rm EG}}{\diff T_\chi\,\diff\Omega}
= \frac{f_{\rm PBH}\,\rho_{\rm DM}}{4\pi\,M_{\rm PBH}}
\int_{t_{\rm min}}^{t_{\rm max}} \diff t\,\big[1+z(t)\big]\,
\frac{\diff^2 N_\chi}{\diff T_\chi\,\diff t}\Big|_{E_\chi^s}\,,
\label{eq:EGnuflux}
\end{eqnarray}
where $\rho_{\rm DM}=2.3\times 10^{-30}~\mathrm{g\,cm^{-3}}$ is the present-day cosmic DM density~\cite{Planck:2018vyg}.
The source DM energy $E_\chi^s$ is related to the observed DM energy $E_\chi$ through cosmological redshift parameter, $z(t)$,
\begin{equation}
E_\chi^s
= \sqrt{\big(E_\chi^2 - m_\chi^2\big)(1+z(t))^2 + m_\chi^2}\,.
\end{equation}
Particles emitted at very early times undergo significant redshifting and fall below detectable energies today. To account for this, we restrict the integration to 
$t_{\rm min} = 10^{11}~\mathrm{s}$, close to matter–radiation equality\footnote{We have explicitly verified that extending the lower limit to earlier times does not lead to any visible change in the resulting extragalactic flux. This is due to the combined effect of suppressed emission rates at early stages of PBH evolution and strong cosmological redshifting, which shifts the emitted particles to energies well below the detection thresholds relevant for our analysis.}~\cite{Wang:2020uvi}, and 
$t_{\rm max} \approx t_{\rm univ}$.
The redshift parameter, $z(t)$, is related to the cosmic time, and in the cosmic regime relevant for our analysis, which predominantly lies in the matter-dominated era, this relation can be well approximated as
\begin{equation}
    z(t) \simeq \left(\frac{2}{3 H_0 \sqrt{\Omega_m}\, t}\right)^{2/3} - 1\,,
\end{equation}
with $H_0\approx68~\mathrm{km\,s^{-1}\,Mpc^{-1}}$ be the Hubble constant~\cite{Planck:2018vyg, SPT-3G:2022hvq}, and $\Omega_m=0.315\pm 0.007$ denotes the present-day matter density parameter~\cite{Planck:2018vyg}, defined as the ratio of the total matter energy density to the critical density of the Universe.

In Fig.~\ref{fig:uan_flux} we present the differential DM flux per unit solid angle emitted from PBH evaporation, $\diff^{2}\Phi_\chi/\diff T_\chi\,\diff\Omega$, as a function of the DM kinetic energy $T_\chi$. The spectra are shown for a DM mass of $m_\chi = 1~\mathrm{MeV}$ and for two benchmark PBH masses, $M_{\rm PBH} = 10^{15}~\mathrm{gm}$ and $10^{16}~\mathrm{gm}$. For each mass we adopt values of the PBH abundance consistent with current cosmological limits \cite{Malyshev:2023oox}, namely $f_{\rm PBH}=3.9\times 10^{-7}$ for $M_{\rm PBH}=10^{15}~\mathrm{gm}$ (red) and $f_{\rm PBH}=5.9\times 10^{-4}$ for $M_{\rm PBH}=10^{16}~\mathrm{gm}$ (blue). The individual contributions to the flux are displayed separately: the Galactic component (dashed), the extragalactic component (dash–dotted), and their sum (solid). 
Comparing the DM fluxes from the aforementioned two source PBHs, we infer that the PBH with mass $M_{\rm PBH}=10^{16}~\mathrm{gm}$ leads to a DM flux almost $\mathcal{O}(3)$ larger than the other one at $T_\chi\lesssim 1$ MeV. 
This feature can be associated primarily with the values of the respective  PBH fractions, $f_{\rm PBH}$.
For $m_\chi+T_\chi\gg T_{\rm PBH}~(\propto 1/M_{\rm PBH})$ the evaporated flux becomes exponentially suppressed (see eq.~\eqref{Eq:DM_number_Spectra}) leading to sharp decrease at higher values of $T_\chi$ in both of the fluxes shown in Fig.~\ref{fig:uan_flux}.
Following the same argument, we find that the flux becomes negligible for $T_\chi \gtrsim 30$ MeV ($300$ MeV) for the source PBH mass $M_{\rm PBH}=10^{16}~\mathrm{gm} ~(10^{15}~\mathrm{gm})$. 
For lower values of $T_\chi~ (\ll m_\chi)$ the DM particles produced are almost at rest and thus the flux become almost independent of $T_\chi$ at that energy range.
Note that the emitted flux of DM is independent of DM-matter interaction type and strength.

\section{Expected recoil signals from DM--electron scattering with realistic cross-sections}\label{Sec:Event_Simulation}
Here, we outline the formalism used to calculate the DM--electron scattering rates and the statistical framework adopted in our analysis, for current direct detection experiments (XENONnT, LZ, and PandaX-4T).
The DM particles $\chi$ emitted via PBH evaporation can subsequently scatter off electrons in terrestrial detectors, producing observable electron recoil signals. 
The expected event rate is given by,
\begin{equation}
    \frac{\diff R}{\diff E_R}=t_{\rm exp} N_{\rm T}Z_{\rm eff}(E_R)\int\diff \Omega\int^{T_{\chi}^{\rm max}}_{T_{\chi}^{\rm min}(E_R)}\diff T_{\chi}\, \frac{\diff^2\Phi_\chi}{\diff T_\chi\diff \Omega} \frac{\diff \sigma_{\chi e}}{\diff E_R}\,,
    \label{eq:recoil}
\end{equation}
where $E_R$ denotes the recoil energy of the target electron, while $t_{\rm exp}$ and $N_{\rm T}$ represent the effective experimental run time and the total number of atoms within the fiducial detector volume, respectively. The effects arising from the binding of atomic electrons in xenon are incorporated through an effective charge function, $Z_{\rm eff}(E_R)$. We model this quantity following the approach of Ref.~\cite{Chen:2016eab}, where the ionization of atomic shells is implemented via a set of step functions. Explicitly, the effective charge is approximated as $Z_{\rm eff}(E_R) \approx \sum_{i=1}^{54} \Theta\!\left(E_R-{\rm BE}_i\right)$, with $\Theta(x)$ denoting the Heaviside step function, and ${\rm BE}_i$ representing the binding energies of the individual electrons in xenon obtained from Hartree--Fock calculations~\cite{Chen:2016eab}. In this framework, following Ref.~\cite{Chen:2016eab}, the scattering kinematics and differential cross-sections are evaluated assuming free electrons,
while $Z_{\rm eff}(E_R)$ ensures that only the electrons having binding energies below the recoil energy contribute to the signal.
$T_{\chi}^{\rm min}$ stands for the  minimum kinetic energy of $\chi$ need to scatter an free electron with a recoil energy $E_R$ and reads as
\cite{Bringmann:2018cvk},
\begin{equation}
    T_{\chi}^{\rm min}(E_R) =
    \left(\frac{E_R}{2}-m_\chi \right) \left[1 \pm \sqrt{1+\frac{2 E_R (m_e+m_\chi)^2}{m_e(2 m_\chi-E_R)^2}} \right] ,
    \label{eq:tmin2}
\end{equation}
with $(+)$ or $(-)$ sign depending on whether $E_R > 2m_\chi$ or $E_R < 2 m_\chi$, respectively.
On the other hand, $T_{\chi}^{\rm max}$ is the maximum kinetic energy of the PBH evaporated DM. In this study we have considered $T_{\chi}^{\rm max}=1$ GeV.
The differential scattering cross-section with the target electron in the heavy mediator limit is given by \cite{Ghosh:2021vkt},
\begin{eqnarray}
    &&{\rm Constant:~}\left(\frac{\diff\sigma_{\chi e}}{\diff E_R}\right)_{\rm c}=\frac{\bar{\sigma}_{\chi e}}{E_R^{\rm max}(T_\chi)},\label{eq:sig_const}
    \end{eqnarray}
for constant cross-section. $\bar{\sigma}_{\chi e}$ is the reference DM--e cross-section with momentum transfer $\alpha m_e$ (where $\alpha$ is Fine structure constant) \cite{Bardhan:2022bdg}.
Here $s_\chi$ is the center of mass energy squared of the DM-e system and $E_R^{\rm max}$ is the maximum recoil energy produced upon scattering with DM with KE $T_\chi$. 
They are given by,
\begin{equation}
    E_R^{\rm max} =\dfrac{T_\chi^2+2 m_\chi T_\chi}{T_\chi+(m_\chi+m_e)^2/(2 m_e)},~~~s_{\chi}=(m_\chi+m_e)^2+2 m_e T_\chi.
    \label{eq:ermax}
\end{equation}

Notably, the constant cross-section approximation is strictly valid only for non-relativistic DM--electron scattering mediated by a heavy particle. In the present context, however, DM particles emitted via PBH evaporation are typically (semi-) relativistic, and their kinetic energies can be comparable to or larger than the electron mass. In this regime, the recoil-energy dependence of the scattering process becomes important and the constant cross-section approximation is no longer reliable. Instead, the differential cross-section acquires a nontrivial dependence on the Lorentz structure of the underlying interaction. In particular, scalar- and vector-mediated interactions lead to distinct recoil spectra emerging from the following form of differential cross-sections,
\begin{eqnarray}
     &&{\rm Scalar~ mediated:~}\left(\frac{\diff\sigma_{\chi e}}{\diff E_R}\right)_{\rm s} = \bar{\sigma}_{\chi e} \left (\dfrac{m_e}{4\mu_{e \chi}^2} \right ) \dfrac{(2m_e+E_R)(2 m_\chi^2+m_e E_R)}{s_{\chi}E_R^{\rm max}},\label{eq:sig_scalar}\\ 
    &&{\rm Vector~ mediated:~}\left(\frac{\diff\sigma_{\chi e}}{\diff E_R}\right)_{\rm v} = \bar{\sigma}_{\chi e} \left (\dfrac{m_e}{2\mu_{e \chi}^2}\right ) \bigg(2 m_e (m_\chi+T_\chi)^2 -E_R ((m_e+m_\chi)^2 \nonumber\\
    &&~~~~~~~~~~~~~~~~~~~~~~~~~~~~~~~~~~~~~~~~~~~~+2 m_e T_\chi )+m_e E_R^2 \bigg)/(s_{\chi}E_R^{\rm max} ),\label{eq:sig_vector}
\end{eqnarray}
for scalar-mediated cross-section, and vector-mediated cross-section, respectively. Motivated by this, in our analysis, we go beyond the constant cross-section approximation and explicitly consider the full relativistic expressions for scalar- and vector-mediated DM--electron scattering when computing recoil rates and deriving experimental constraints.

\begin{figure}[!tbh]
    \centering
    \subfigure[\label{xeu}]{
    \includegraphics[scale=0.5]{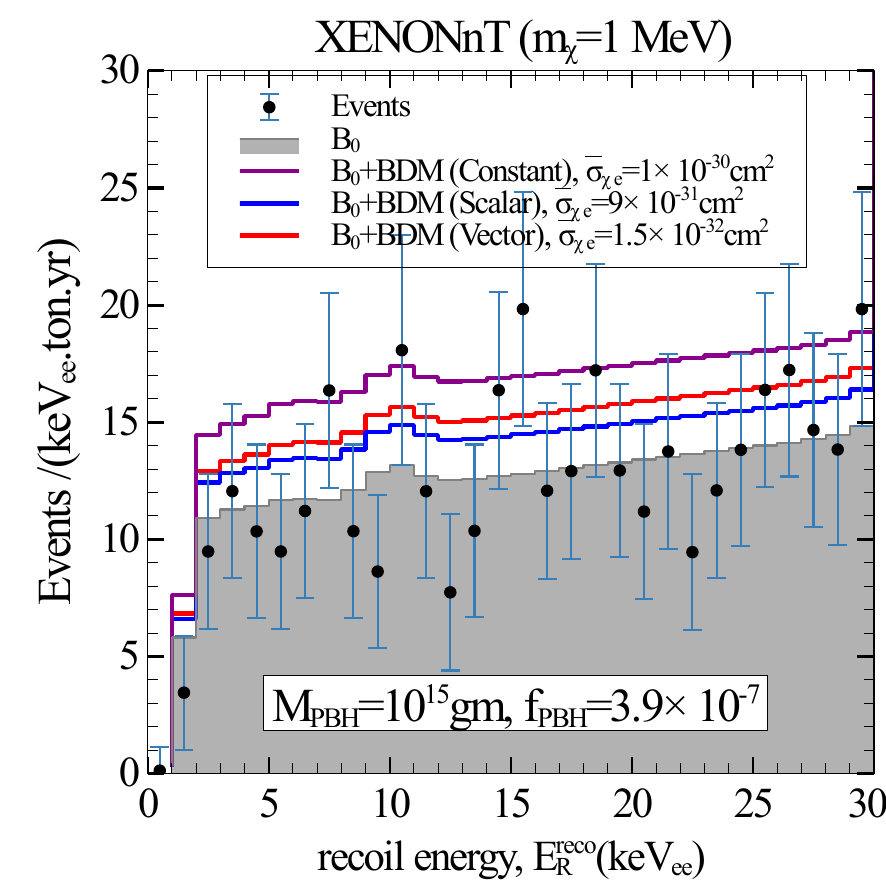}}
    \subfigure[\label{Lzu}]{
    \includegraphics[scale=0.5]{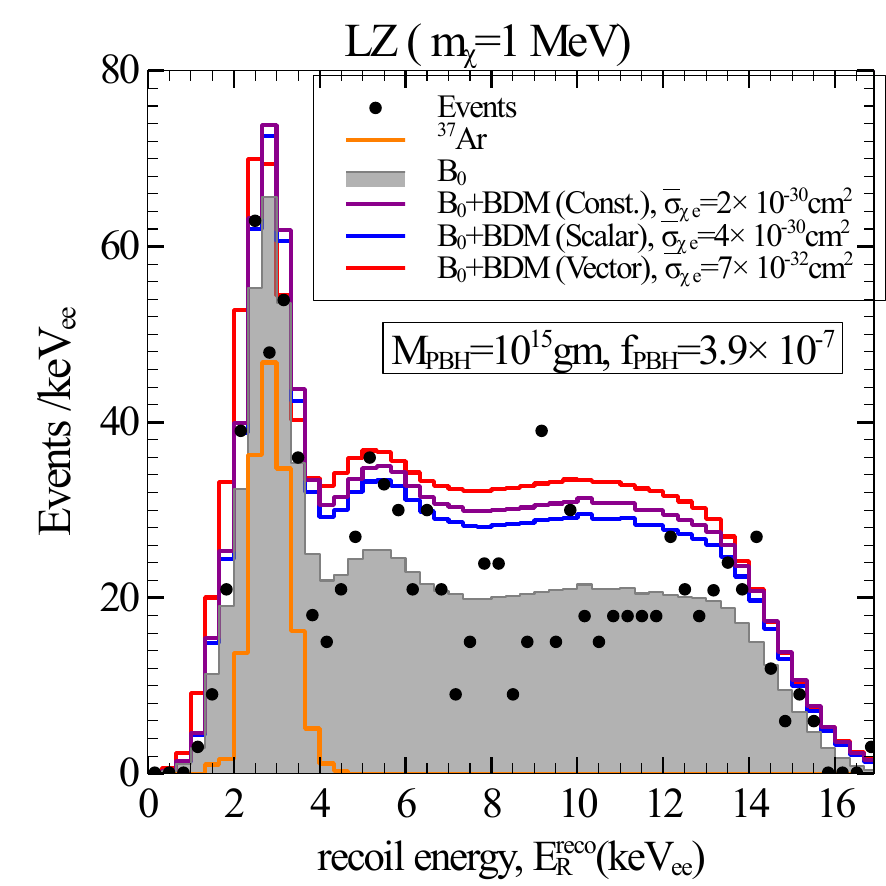}}
    \subfigure[\label{pdu}]{
    \includegraphics[scale=0.5]{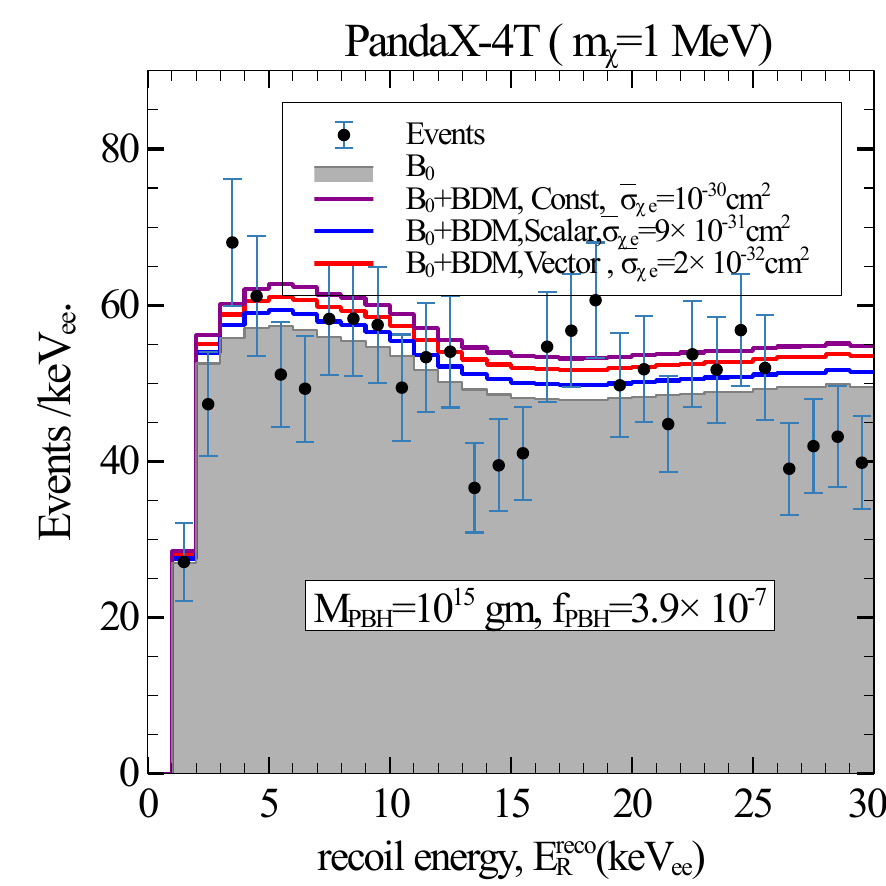}}
    \caption{The event spectra of $e$ upon being scattered by $\chi$ evaporated from a PBH with $M_{\rm PBH}=10^{15}$ gm and $f_{\rm PBH}=3.9 \times 10^{-7}$ in (a) XENONnT, (b) LZ and (c) PandaX-4T.
    Rates for different cross-sections e.g constant (magenta), vector (red) and scalar (blue) are depicted by different solid colored lines. Grey region signifies the known backgrounds denoted as $B_0$ where as the black dots with (without in (b)) error bars signify the observed event data.}
    \label{fig:event_uan}
\end{figure}
We stress that the energy-dependent cross-sections are written in the limit when the associated mediators (scalar/vector) are heavier than the associated DM momentum transfer.
However, any BSM mediator present in the particle content is also expected to evaporate from the PBH and may potentially decay into DM, thereby acting as a secondary source of DM and enhancing the overall flux. 
Note that, we assume the simpler scenario where the mediators are much heavier than $T_{\rm PBH}$, such that their Hawking emission is exponentially suppressed and hence effectively negligible.
In principle, as the PBH evaporates, its temperature increases, allowing additional particle species to be emitted at late times.
However, for the range of PBH masses considered in this work, they are long-lived and survive till today, ensuring our assumption of the mediator mass remains valid. 
Under this assumption, the mediators neither give rise to any appreciable contribution to the Hawking emission nor significantly alter the particle spectrum, allowing us to consistently describe the DM--electron interaction within an effective field theory framework \cite{Calabrese:2022rfa}.


To compute the event spectra, we then plug the aforementioned cross-sections in eq.~\eqref{eq:recoil}.
Finally the reconstructed event spectra is obtained after using a smearing given as,
\begin{eqnarray}
    \frac{\diff R}{\diff E_R^{\rm reco}} =\mathcal{E}(E_{R}^{\rm reco}) \int_{0}^{E_R^{\rm max}} dE_R ~\frac{\diff R}{\diff E_R} \mathcal{G}(E_{R},E_{R}^{\rm reco}).
\end{eqnarray}
$E_{R}^{\rm reco}$ is the reconstructed electron recoil energy and $\mathcal{G}(E_{R},E_{R}^{\rm reco})$ is the Gaussian smearing function which reads as,
\begin{eqnarray}
\label{Eq:Smeared_Spectra}
    \mathcal{G}(E_{R},E_{R}^{\rm reco}) =\frac{1}{\sqrt{2\pi}\sigma}\exp{\left(-\dfrac{(E_{R} -E_{R}^{\rm reco})^2}{2\sigma^2} \right)}.
\end{eqnarray}
$\sigma$ is the resolution power given by~\cite{XENON:2020iwh, Pereira:2023rte, PandaX:2022ood},
\begin{eqnarray}
    {\rm XENONnT:~}&& \sigma=0.31 \sqrt{E_{R}^{\rm reco}}+0.0037 E_{R}^{\rm reco}\\
    {\rm LZ:~}&& \sigma=0.323\sqrt{E_{R}^{\rm reco}}\\
    {\rm PandaX-4T:~}&& \sigma=0.073+0.173 E_{R}-6.5\times 10^{-3} E_R^2+1.1\times 10^{-4}E_{R}^{3}
\end{eqnarray}
The function $\mathcal{E}(E_{R}^{\rm reco})$ in eq.~\eqref{Eq:Smeared_Spectra} represents the recoil-energy-dependent detection efficiency. In our analysis, the efficiency curves for XENONnT, LZ, and PandaX-4T are taken directly from Refs.~\cite{XENON:2022ltv}, \cite{LZ:2023poo}, \cite{PandaX:2024cic} respectively.

The expected event spectra is shown in Fig.~\ref{fig:event_uan}. 
We showcase the expected recoil signature  for a fixed mass $m_\chi =1$ MeV in XENONnT (Fig.~\ref{xeu}), LZ (Fig.~\ref{Lzu}) and PandaX-4T (Fig.~\ref{pdu}).
For the source PBH we consider $M_{\rm PBH}=10^{15}$ gm and PBH fraction $f_{PBH}=3.9\times 10^{-7}$ which allowed from the existing constraints on PBH \cite{Malyshev:2023oox}.
The envelope of the grey region signifies the known backgrounds denoted as $B_0$ where as the 
black dots with (without in Fig.~\ref{Lzu}) error bars signify the observed event data.
The magenta, blue and red lines signify the recoil rates for $\chi$ with constant, scalar and vector type of interaction.
For XENONnT in Fig.~\ref{xeu} we consider $\bar{\sigma}_{\chi e}=  10^{-30}$cm$^2$ (magenta), $\bar{\sigma}_{\chi e}= 9\times 10^{-31}$cm$^2$ (blue) and $\bar{\sigma}_{\chi e}= 1.5\times 10^{-32}$cm$^2$ (red) for constant, scalar and vector cross-sections respectively.
For LZ in Fig.~\ref{Lzu} we consider $\bar{\sigma}_{\chi e}= 2\times 10^{-30}$cm$^2$ (magenta), $\bar{\sigma}_{\chi e}= 4\times 10^{-30}$cm$^2$ (blue) and $\bar{\sigma}_{\chi e}= 7\times 10^{-32}$cm$^2$ (red) for constant, scalar and vector cross-sections respectively.
Finally for PandaX-4T in Fig.~\ref{pdu}we choose  $\bar{\sigma}_{\chi e}=  10^{-30}$cm$^2$ (magenta), $\bar{\sigma}_{\chi e}= 9\times 10^{-31}$cm$^2$ (blue) and $\bar{\sigma}_{\chi e}= 2\times 10^{-32}$cm$^2$ (red) for constant, scalar and vector cross-sections respectively.
Note that $\chi$ with different interaction types e.g. constant, scalar and vector, lead to almost similar order of expected signals in each detector, with quite different cross-section strength.
This feature is evident in all of the aforementioned plots and serves as the primary motivation for this study to investigate various interactions of $\chi$.

After having a detailed discussion of the procedure for computing the recoil spectra induced by PBH-evaporated dark particles at direct detection experiments, we are now set to perform the parameter scan to place constraints on the $m_\chi$ vs. $\bar{\sigma}_{\chi e}$ plane. For that we first define the statistical framework and the corresponding test statistic employed to analyze the direct detection data. In constructing our statistical framework, we follow the approach outlined in Ref.~\cite{Majumdar:2024dms} (see also Ref.~\cite{A:2022acy,DeRomeri:2023ytt}).

For XENONnT and PandaX-4T we perform a binned $\chi^2$ analysis, incorporating a single nuisance parameter to account for background normalization uncertainties. For this case, we adopt a Gaussian $\chi^2$ test statistic, defined as\footnote{We have verified that adopting a Poissonian $\chi^2$ treatment, analogous to eq.~\eqref{eq:chi2_LZ}, for XENONnT and PandaX-4T does not lead to any visible change in the resulting parameter space.}~\cite{Almeida:1999ie},
\begin{eqnarray}
\label{eq:chi2_XENON_Panda}
    \chi^2(\overrightarrow{\mathcal{S}};\alpha)
    = \sum_{i=1}^{N_{\rm bin}} 
    \left( 
    \frac{R_{\rm pred}^i(\overrightarrow{\mathcal{S}};\alpha)-R_{\rm exp}^i}
    {\sigma^i}
    \right)^2
    + \left(\frac{\alpha}{\sigma_\alpha}\right)^2 ,
\end{eqnarray}
$\overrightarrow{\mathcal{S}}$ signifies the set of BSM parameters i.e. the $2-$dimension space of $m_\chi, \bar{\sigma}_{\chi e}$.
$R_{\rm pred}^i$ is the predicted event in $i$-th bin and for a fixed parameter point it is defined as,
\begin{equation}
R_{\rm pred}^i(\overrightarrow{\mathcal{S}};\alpha)
=(1+\alpha)R_{\rm bkg}^i+R_{\rm BSM}^i(\overrightarrow{\mathcal{S}})\,.
\end{equation}
$R_{\rm exp}^i$ and $\sigma_i$ are the experimentally observed data and associated statistical uncertainty in $i$-th bin.
For XENONnT, the experimental data and the modeled background $B_0$ are taken from Ref.~\cite{XENON:2022ltv}, while for PandaX-4T (Run 1) we use the publicly reported background and data from Ref.~\cite{PandaX:2024cic}.
In the case of XENONnT, the collaboration provides a best-fit background model $B_0$ obtained from their dedicated likelihood analysis; we therefore do not include an additional normalization nuisance parameter $\alpha$ in eq.~\eqref{eq:chi2_XENON_Panda} for this dataset. By contrast, for PandaX-4T (Run~1) we retain the nuisance parameter $\alpha$ in eq.~\eqref{eq:chi2_XENON_Panda} to account for background uncertainties, adopting $\sigma_\alpha = 2.5\%$ in our analysis.

For LZ detector we follow Possionian $\chi^2$ analysis~\cite{Almeida:1999ie},
\begin{eqnarray}
\label{eq:chi2_LZ}
    \chi^2(\overrightarrow{\mathcal{S}}; \alpha,\delta)&=&2\sum_{i=1}^{51} \Bigg[  R_\mathrm{pred}^i(\overrightarrow{\mathcal{S}}; \alpha,\delta) - R_\mathrm{exp}^i+  R_\mathrm{exp}^i\ln \left(\frac{R_\mathrm{exp}^i}{ R_\mathrm{pred}^i(\overrightarrow{\mathcal{S}}; \alpha,\delta)}\right) \Bigg] \nonumber \\
    &&+\left(\frac{\alpha}{\sigma_\alpha}\right)^2+\left(\frac{\delta}{\sigma_\delta}\right)^2 ,
\end{eqnarray}
For LZ, the experimental data and background components are taken from Ref.~\cite{LZ:2022lsv}, while the total predicted events in each recoil bin are calculated using
\begin{equation}
\label{eq:LZ_R_pred}
R_{\rm pred}^i(\overrightarrow{\mathcal{S}}; \alpha,\delta)=(1+\alpha)R_{\rm bkg}^i+(1+\delta)R_{^{37}{\rm Ar}}^i+R_{\rm BSM}^i(\overrightarrow{\mathcal{S}})\,.
\end{equation}
where $\alpha$ and $\delta$ are the nuisance parameters accounting for the uncertainty in the total background and in $^{37}Ar$ component of the background with $\sigma_\alpha=13\% $ and $\sigma_\delta=100\%$. Following Ref.~\cite{A:2022acy}, the background spectrum $R_{\rm bkg}^i$ is obtained by subtracting the $^{37}$Ar contribution from the total background reported by the LZ Collaboration~\cite{LZ:2022lsv}, since in eq.~\eqref{eq:LZ_R_pred} the uncertainty associated with the $^{37}$Ar component is treated separately through the nuisance parameter $\delta$.

Finally for each set of BSM parameters, we marginalize the $\chi^2$ functions over all nuisance parameters discussed above and  obtain the $\Delta\chi^2=\chi^2-\chi^2_{\rm min}$. From that we place the $2\sigma$ confidence level (C.L) exclusion limit on the parameter space.

\section{Results without attenuation effects}
\label{sec:result_unatt}

In this section, we present the constraints on PBH-evaporated DM obtained under the assumption that the DM flux reaches terrestrial detectors without undergoing any in-medium attenuation. This unattenuated scenario serves as a useful baseline to illustrate the intrinsic sensitivity of direct detection experiments to energetic DM particles produced via PBH evaporation. We focus on the resulting limits in the DM mass and DM--electron scattering cross-section plane for different interaction structures and benchmark PBH parameters.

\begin{figure}[ht!]
    \centering
    \subfigure[\label{cu1}]{
    \includegraphics[scale=0.5]{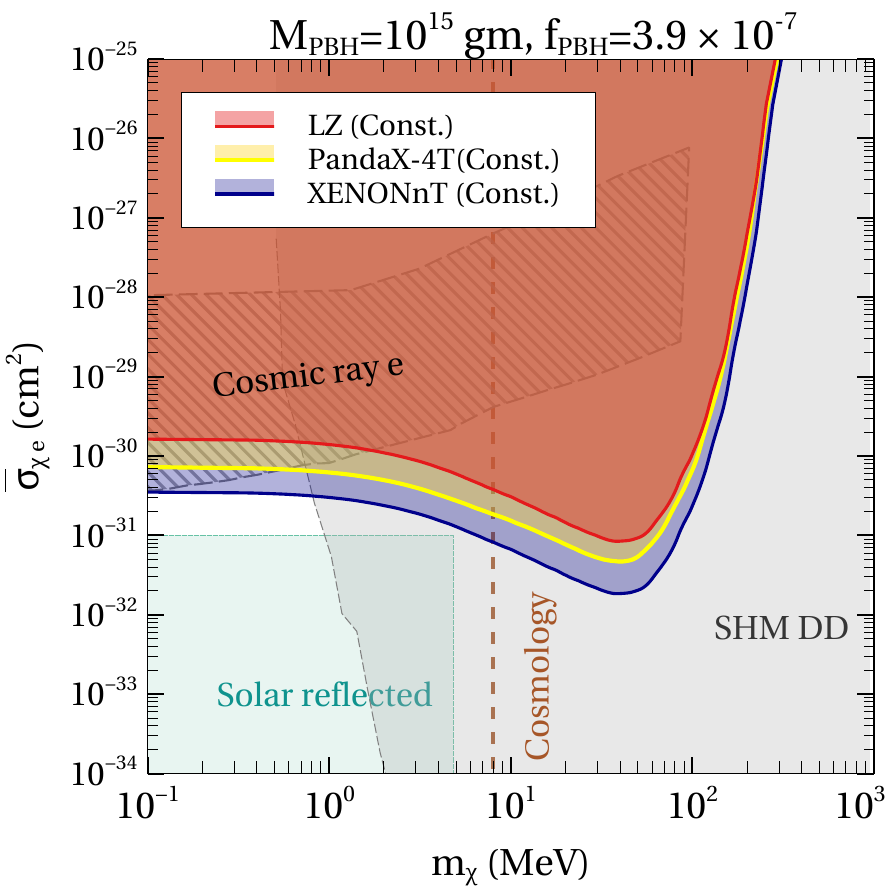}}
    \subfigure[]{
    \includegraphics[scale=0.5]{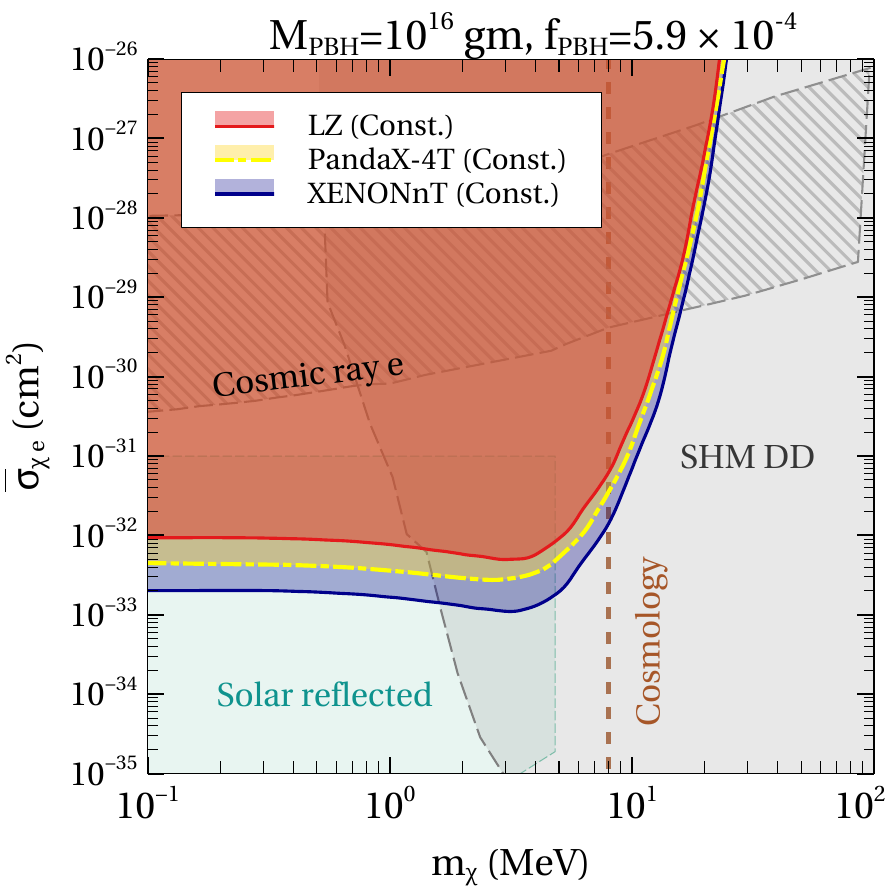}}
    \caption{{\bf Constant cross-section:} Constraints at $2\sigma$ C.L. on DM $\chi$ with {\bf constant} cross-section evaporated from PBH  using XENONnT (light blue), PandaX-4T (light yellow) and LZ (light red) . Chosen PBH BPs are (a) $M_{\rm PBH}=10^{15}$ gm with $f_{\rm PBH}=3.9\times 10^{-7}$ and (b) $M_{\rm PBH}=10^{16}$ gm with $f_{\rm PBH}=5.9\times 10^{-4}$. Other existing constraints are also displayed.  }
    \label{fig:con_u}
\end{figure}

\begin{figure}[ht!]
    \centering
    \subfigure[\label{vu1}]{
    \includegraphics[scale=0.5]{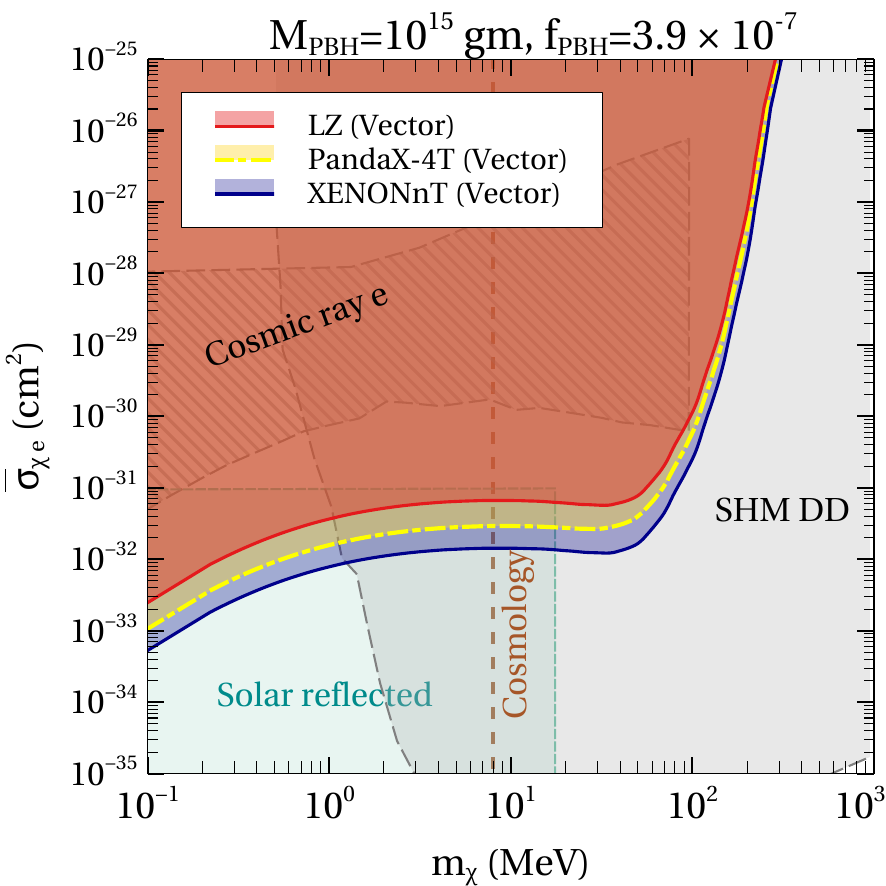}}
    \subfigure[]{
    \includegraphics[scale=0.5]{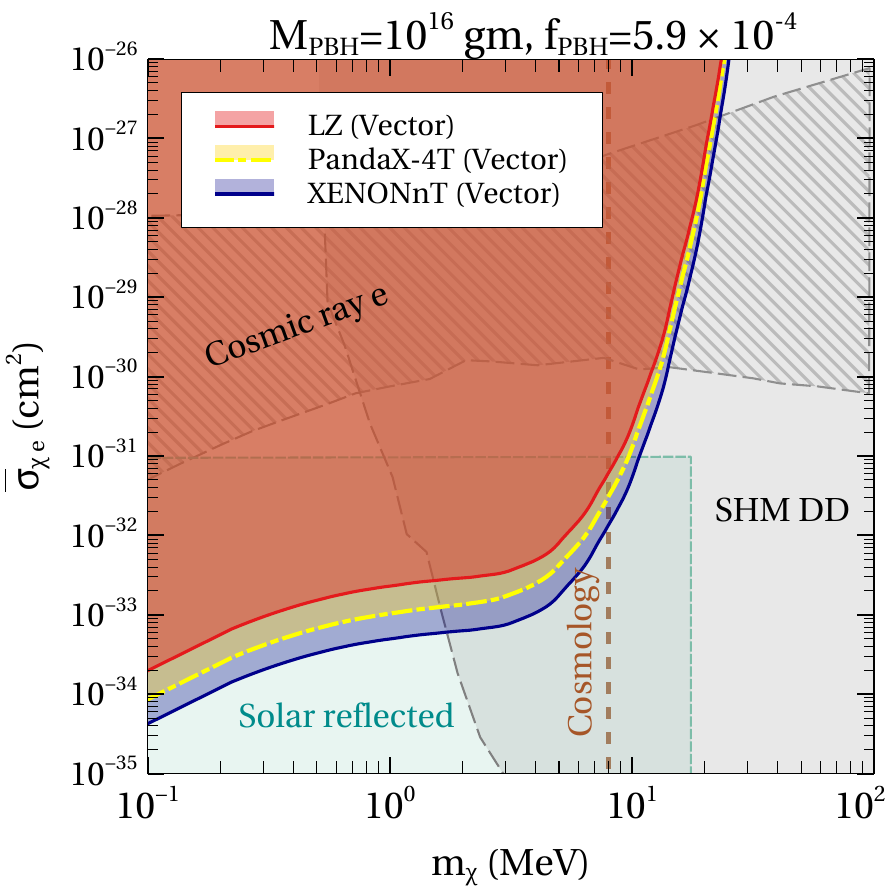}}
    \caption{ {\bf Vector cross-section:} Constraints  at $2\sigma$ C.L. on DM $\chi$ with {\bf vector} cross-section evaporated from PBH  using XENONnT (light blue), PandaX-4T (light yellow) and LZ (light red) . Chosen PBH BPs are (a) $M_{\rm PBH}=10^{15}$ gm with $f_{\rm PBH}=3.9\times 10^{-7}$ and (b) $M_{\rm PBH}=10^{16}$ gm with $f_{\rm PBH}=5.9\times 10^{-4}$. Other existing constraints are also displayed.}
    \label{fig:vec_u}
\end{figure}

\begin{figure}[ht!]
    \centering
    \subfigure[\label{su1}]{
    \includegraphics[scale=0.5]{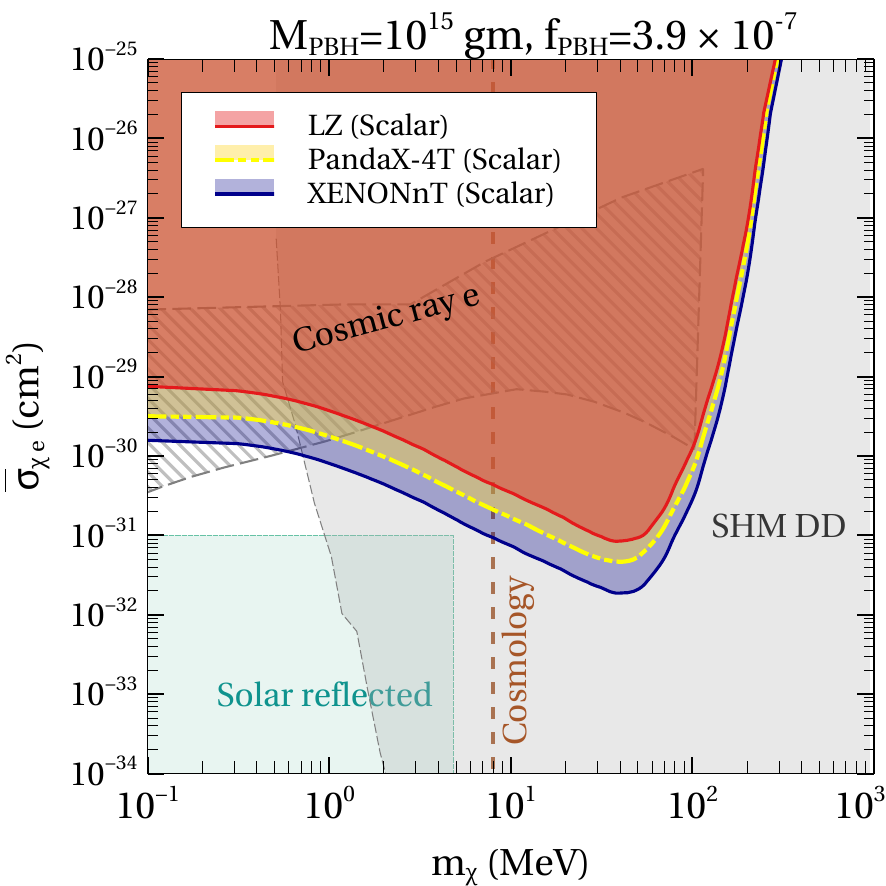}}
    \subfigure[]{
    \includegraphics[scale=0.5]{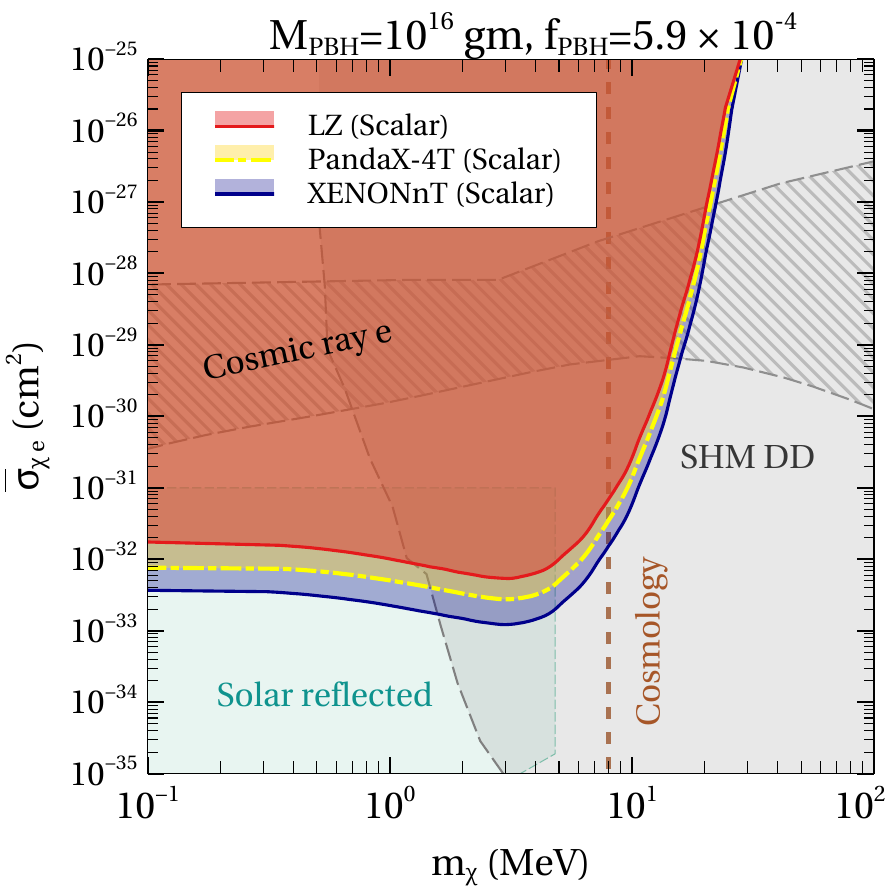}}
    \caption{{\bf Scalar cross-section:} Constraints  at $2\sigma$ C.L. on DM $\chi$ with {\bf scalar} cross-section evaporated from PBH  using XENONnT (light blue), PandaX-4T (light yellow) and LZ (light red) . Chosen PBH BPs are (a) $M_{\rm PBH}=10^{15}$ gm with $f_{\rm PBH}=3.9\times 10^{-7}$ and (b) $M_{\rm PBH}=10^{16}$ gm with $f_{\rm PBH}=5.9\times 10^{-4}$. Other existing constraints are also displayed.}
    \label{fig:scal_u}
\end{figure}
In Fig.~\ref{fig:con_u}, \ref{fig:vec_u} and \ref{fig:scal_u} we present our obtained constraints on PBH evaporated $\chi$ from DM detectors. We showcase our limits in the $m_\chi$ vs. $\bar{\sigma}_{\chi e}$ for different types of cross-sections i.e. constant, vector and scalar mediated, respectively. 
As hinted earlier, we choose PBH as the source of $\chi$ with the following two BP parameters: (1)$M_{\rm PBH}=10^{15}$ gm with $f_{\rm PBH}=3.9\times 10^{-7}$ ({\bf left panels}) and (2) $M_{\rm PBH}=10^{16}$ gm  with $f_{\rm PBH}=5.9\times 10^{-4}$ ({\bf right panels}) allowed from the existing limits \cite{Malyshev:2023oox}. Constraints at $2\sigma$ confidence limit (C.L.) from XENONnT, LZ and PandaX-4T are indicated by light blue, light red and light yellow shaded regions. In all of the aforementioned plots XENONnT places the most stringent limit as expected from the larger exposure of it compared to PandaX-4T and LZ.

For a fixed $M_{\rm PBH}$, different cross-section types lead to significantly different constraints from DM detectors as observed in the aforementioned plots.
For e.g. comparing the {\bf left columns} i.e. Fig.~\ref{cu1},~\ref{vu1} and \ref{su1} we find that, 
vector one leads to most stringent limit, followed by constant and scalar ones.
The vector-mediated cross-section leads to limits stronger ($\mathcal{O}(3)$) than those with a constant cross-section. 
On the other hand limits obtained with scalar mediated cross-sections are weaker by factor $\sim6$ than those with constant cross-section.
A similar conclusion also holds for DM evaporated from other source PBHs.
Thus we justify the importance of considering energy-dependent cross-sections while studying the direct searches of PBH evaporated DM.
Comparing the left and right panels each figure (Fig.~\ref{fig:con_u}, \ref{fig:vec_u} and \ref{fig:scal_u}) it can be apprehended that, in the  MeV mass range, $\chi$ evaporated from a PBH with $M_{\rm PBH}=10^{16}$ gm face stronger constraint for all 3 cross-sections. This feature can be understood from the higher fraction $f_{\rm PBH}=5.9\times 10^{-4}$ compared to the other one.
However, the constraints become significantly weaker for $m_\chi\gtrsim \mathcal{O}$(100) MeV ($\mathcal{O}$(10) MeV)  for the PBH BP with $M_{\rm PBH}=10^{15}$ gm ($M_{\rm PBH}=10^{16}$ gm).
This can be associated with the maximum energy a $\chi$ particle can have after getting produced from a PBH (see Fig.~\ref{fig:uan_flux}).

In the same planes of Fig.~\ref{fig:con_u}, \ref{fig:vec_u} and \ref{fig:scal_u} we also display other existing constraints on boosted light DM that could be relevant for our analysis. Examples of such limits include the cosmic electron boosted DM in XENON1T \cite{Bardhan:2022bdg} (grey hatched region) and solar reflected DM \cite{An:2017ojc,Emken:2024nox} (light cyan region).
The combined limit for galactic halo DM assuming standard halo model (SHM) from SuperCDMS \cite{SuperCDMS:2018mne}, SENSEI \cite{Crisler:2018gci}, and XENON10 \cite{Essig:2012yx} is shown by the grey shaded region.
We also portray the cosmological bound on fermionic DM from BBN \cite{Krnjaic:2019dzc} shown by brown dashed line throughout the draft.
However, this limit depends on the thermal history of $\chi$. 
At this point we stress that, if one assumes $\chi$ to be a generic long-lived particle, then these DM constraints do not apply. 
Thus, DM detectors can be a useful probe of such dark sector particles evaporated from PBH.

\section{Role of attenuation}\label{Sec:Attenuation_Technique}

So far, we do not consider any possible attenuation effect on the incoming $\chi$ flux. 
Since the cross-sections we are dealing with are quite higher (compared to GeV scale DM searches), DM might scatter significantly with the ambient $e$ while traversing through the atmosphere and Earth.  In this work, however, we neglect attenuation in the atmosphere, as the atmospheric electron density is several orders of magnitude smaller than that of the Earth's interior~\cite{DeRomeri:2023ytt}, rendering the corresponding energy loss negligible for the range of cross-sections considered here.

The inclusion of Earth's attenuation will inevitably change the incoming flux of $\chi$ and hence the constraints from the underground detectors.
The KE of $\chi$ after traveling a distance $z$ through Earth is given by the differential energy loss equation \cite{Bringmann:2018cvk,Xia:2021vbz}, 
\begin{eqnarray}
    \frac{\diff T_\chi^z}{\diff z} = - n_e \int_{0}^{T_e^{\rm max}(T_\chi^z)} \frac{\diff \sigma_{\chi e}}{\diff T_e}~ T_e~ dT_e,
    \label{eq:attn}
\end{eqnarray}
where \cite{DeRomeri:2023ytt},
\begin{eqnarray}
    z=-(R_{\oplus}-h_d)\cos{\theta_z}+\sqrt{R_{\oplus}^2-(R_\oplus-h_d)^2\sin^2{\theta_z}}.
    \label{eq:z}
\end{eqnarray}
$T_e^{\rm max}(T_\chi^z)$ is the maximum KE of the outgoing $e$ after being scattered by $\chi$ with KE $T_\chi^z$ and is given by the same equation of $E_R^{\rm max}$ in eq.~\eqref{eq:ermax} with the replacement $T_\chi\to T_\chi^z$. 
On the other hand, $\theta_z$ is the zenith angle of the detector, while $R_{\oplus}$ and $h_d$ are Earth's radius and the depth of the detector at zero zenith angle, respectively. In our analysis, we take $h_d = 1.4~\mathrm{km}$ for XENONnT, $h_d = 1.47~\mathrm{km}$ for LZ, and $h_d = 2.4~\mathrm{km}$ for PandaX-4T.
eq.~\eqref{eq:attn} is a differential equation with the initial condition $T_\chi^z(z=0)=T_\chi^0$, which is the KE of $\chi$ before entering the Earth's crust. $n_e$ signifies the average electron number density within the Earth. In this work we have considered $n_e=8\times 10^{23}~{\rm cm}^{-3}$~\cite{DeRomeri:2023ytt}.
Finally the flux at the detector is given by~\cite{Bringmann:2018cvk,DeRomeri:2023ytt},
\begin{equation}
\frac{\diff \Phi_\chi}{\diff T_\chi^z}=\int \diff \Omega \left.\frac{\diff ^2\Phi^{\rm PBH}_\chi}{\diff T_\chi \diff\Omega}\right|_{T_\chi^0} \frac{\diff T_\chi^0}{\diff T_\chi^z} ,
\label{eqn:Flux_att}
\end{equation}
where, $\left.\frac{\diff ^2\Phi^{\rm PBH}_\chi}{\diff T_\chi \diff\Omega}\right|_{T_\chi^0}$ signifies the incoming flux of $\chi$ with KE ${T_\chi^0}$ and is obtained as mentioned in Sec.~\ref{Sec:PBH_Flux}.

\subsection{Analytical treatment}
 An approximate solution of the previously mentioned energy loss equation (eq.~\eqref{eq:attn}) is considered in most of the existing literature \cite{Bringmann:2018cvk,Ghosh:2021vkt,Li:2022jxo,Calabrese:2021src,Calabrese:2022rfa, DeRomeri:2023ytt}   which follows as,
\begin{equation}
    T_\chi^z\approx\frac{T_\chi^0e^{-z/l_E}}{1+\frac{T_\chi^0}{2m_\chi}\left(1-e^{-z/l_E}\right)},
    \label{eqn:Txz_e}
\end{equation}
where $l_E$ depicts the mean free path of $\chi$ given by,
\begin{eqnarray}
    l_E^{-1}=n_e\bar{\sigma}_{\chi e}\frac{2m_em_\chi}{(m_e+m_\chi)^2}.
\end{eqnarray} 
In the aforementioned equation, it is assumed that $(m_\chi+m_e)^2\ll 2 m_e T_\chi^z$.
From eq.~\eqref{eqn:Txz_e} one can also obtain $T_\chi^0$ as a function of $T_\chi^z$ and $z$ which looks as,
\begin{equation}
    T_\chi^0\approx \frac{2m_\chi T_\chi^z e^{z/l_E}}{2m_\chi+T_\chi^z\left(1-e^{z/l_E}\right)} .
\end{equation}
Thus the attenuated DM flux in the Earth based detectors can be evaluated simply as,
\begin{equation}
    \frac{\diff \Phi^{\rm Ana.}_\chi}{\diff T_\chi^z}\approx \int \diff\Omega \left.\frac{\diff^2\Phi^{\rm PBH}_\chi}{\diff T_\chi \diff\Omega}\right|_{T_\chi^0} \frac{4m_\chi^2e^{z/l_E}}{\left[2m_\chi+T_\chi^z\left(1-e^{z/l_E}\right)\right]^2}
    \label{eq:flux_analy}
\end{equation}
However, for boosted DM this solution is not valid throughout the DM mass region and is also not applicable for the energy-dependent interactions \cite{Das:2024ghw}.
Previous works considering PBH evaporated DM flux have also derived their limits from DD experiments using such analytic treatment of attenuation effect \cite{Li:2022jxo,Calabrese:2021src,Calabrese:2022rfa}.
We show that the flux of $\chi$ differs significantly after considering the attenuation effect numerically, which stands as the other major thrust of this analysis.
Some of the existing literature also put an upper limit on $\bar{\sigma}_{\chi e}$ above which $\chi$ will have too low $T_\chi$ to produce any electron recoil with energy $> E_{th}$, threshold of the experiment. However, attenuation has a more crucial effect than just setting a ceiling of the constraint plot, as we will see in this section.

\subsection{Numerical treatment}

\begin{figure}[!tbh]
    \centering
    \subfigure[\label{c1}]{
    \includegraphics[scale=0.35]{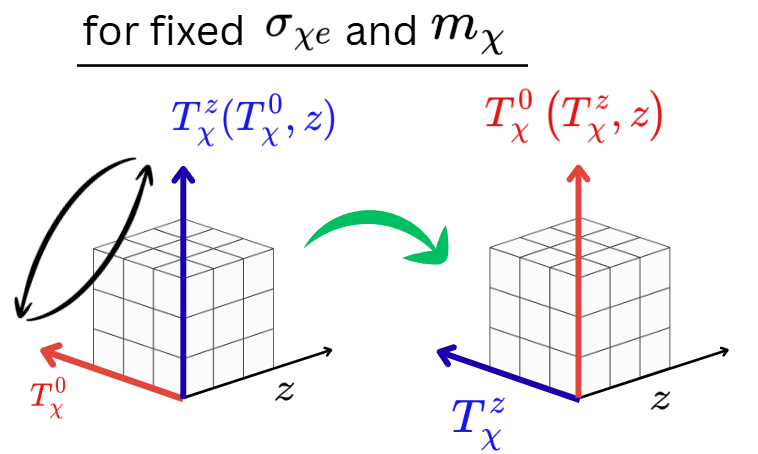}}~~~~~~~~~
    \subfigure[\label{c2}]{
    \includegraphics[scale=0.15]{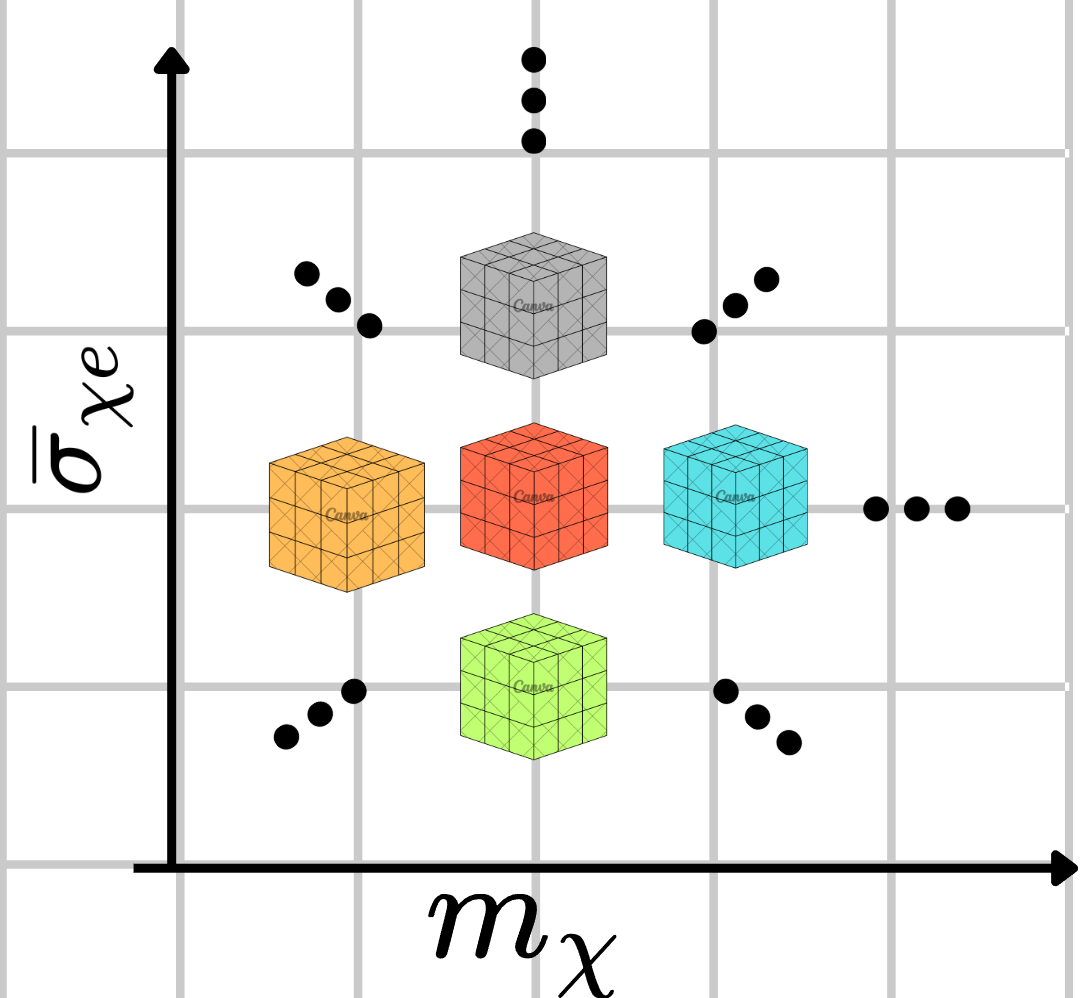}}
    \caption{Pictorial representation of the methodology that we follow to numerically evaluate the attenuated flux of $\chi$ at distance $z$.}
    \label{fig:placeholder}
\end{figure}

To obtain the numerical solution and for full parameter space scan we go along with the following steps-
\begin{enumerate}
    \item For a given $m_\chi$ and $\bar{\sigma}_{\chi e}$ we numerically solve eq.~\eqref{eq:attn} to obtain $T_\chi^z$ at a distance $z$ with initial KE $T_\chi^0$. 
    The range of $T_\chi^0$ is chosen from $1$ keV to $1$ GeV as relevant for our analysis.
    Thus we can find $T_\chi^z$ as function of $z$ and $T_\chi^0$ i.e. $T_\chi^z (z,T_\chi^0)$ for a whole grid of $\{z,T_\chi^0\}$ (See the left grid in Fig.~\ref{c1}). The range of $z$ is chosen from substituting the zenith angle $\theta\in[0,\pi]$ and $h_d$ (depends on specific detector) in eq.~\eqref{eq:z}. 
    \item In the next step we simply flip the cubic grid to find $T_\chi^0 (z,T_\chi^z)$ so that we can interpolate to obtain the value $T_\chi^0$ required to generate $T_\chi^z$ at a given $z$ (See the right grid in Fig.~\ref{c1}). From the same grid we evaluate the Jacobian $\frac{\diff T_\chi^0}{\diff T_\chi^z}$ which we substitute in eq.~\eqref{eqn:Flux_att} and integrate over the zenith angle to obtain the actual (differential)  flux of $\chi$ at the underground detector.
    \item After obtaining the attenuated flux of $\chi$ we follow the same methodology as described in Sec.~\ref{Sec:Event_Simulation} to obtain the recoil signature and then perform a numerical scan as in  Sec.~\ref{sec:result_unatt} to obtain the constraints in the $m_\chi$ vs. $\bar{\sigma}_{\chi e}$ parameter space.
    For the parameter space scan we follow the same first two steps mentioned above on the $2-$D grid of $m_\chi$ vs. $\bar{\sigma}_{\chi e}$ plane (See Fig.~\ref{c2}).
\end{enumerate}
We execute each of these three steps one by one in the following discussion.
\begin{figure}[ht!]
    \centering
    \includegraphics[scale=0.6]{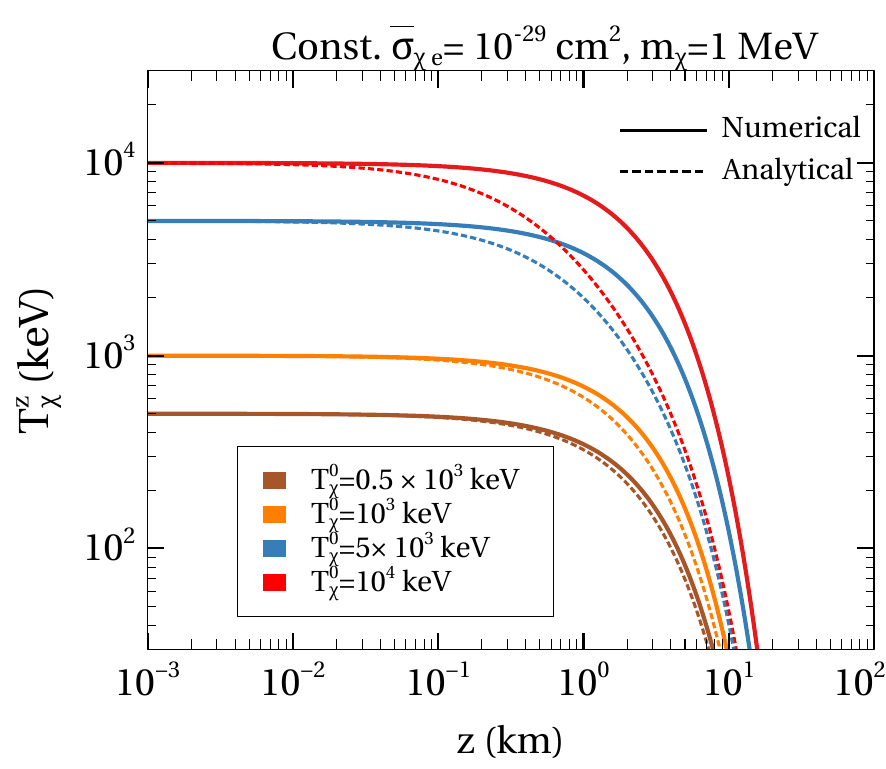}
    \caption{{\bf Step 1.} Comparison between numerical (solid lines) and analytical (dashed lines) solutions of the attenuation equation for the DM kinetic energy $T_\chi^z$ as a function of the propagation distance $z$. The results are shown for $m_\chi=1~\mathrm{MeV}$ and a constant, energy-independent DM--electron cross-section $\bar{\sigma}_{\chi e}=10^{-29}~\mathrm{cm}^2$, for several benchmark initial energies $T_\chi^0$.}
    \label{fig:nu_vs_ana}
\end{figure}

{\bf Step 1.} Following the first step mentioned above we obtain the numerical solution of eq.~\eqref{eq:attn} to get $T_\chi^z$ at a distance $z$. 
In Fig.~\ref{fig:nu_vs_ana} we compare the numerical (solid lines) and analytical (dashed lines) solutions of the attenuation equation for the propagation of PBH-evaporated DM through the Earth. For illustration, we consider a benchmark DM mass of $m_\chi = 1~\mathrm{MeV}$ and an energy-independent constant DM--electron cross-section $\bar{\sigma}_{\chi e}=10^{-29}~\mathrm{cm}^2$. The figure shows the DM kinetic energy $T_\chi^z$ as a function of the propagation distance $z$ for several representative initial energies $T_\chi^0\in \{5\times10^2~{\rm keV},~10^3~{\rm keV},~5\times10^3~{\rm keV},~10^4~{\rm keV} \}$ denoted by brown, orange, light blue and red colors respectively. At shallow depths the numerical and analytical treatments agree reasonably well; however, for $T_\chi^0 \gtrsim 1~\mathrm{MeV}$ the two approaches begin to diverge significantly at larger $z$. In particular, the analytical approximation systematically overestimates the energy loss compared to the full numerical solution, leading to appreciable differences in the predicted $T_\chi^z$ deep inside the Earth. This demonstrates that for DM produced by PBHs, a numerical treatment of attenuation is essential for reliable predictions, especially for large propagation distances through the Earth.

\begin{figure}[ht!]
    \centering
    \includegraphics[scale=0.6]{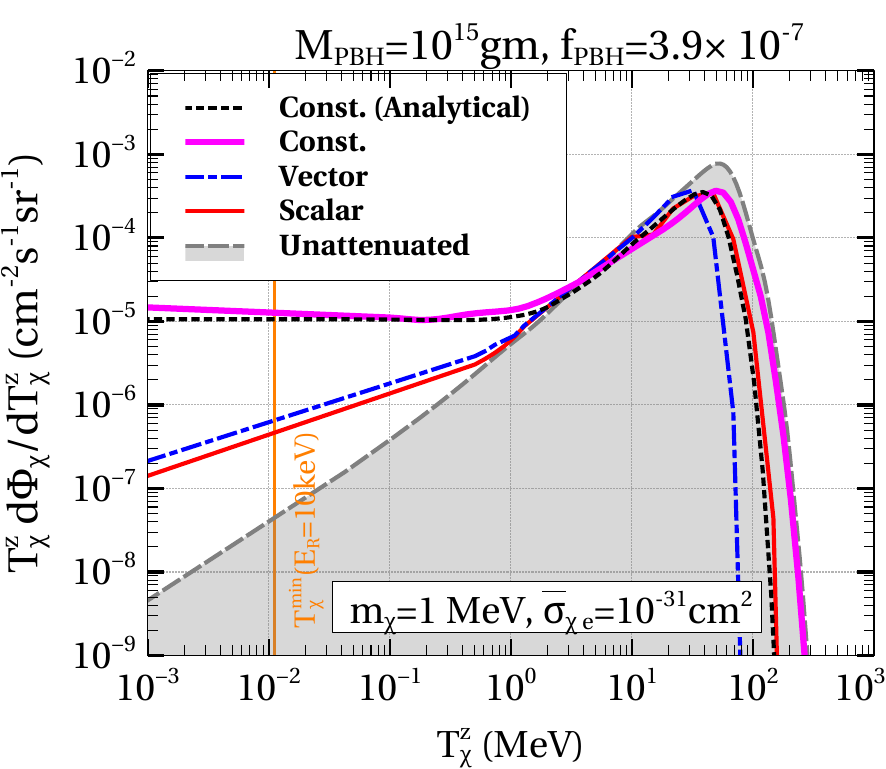}
    \caption{{\bf Step 2.} Variation of differential flux weighted by $T_\chi^z$ with respect to KE of $\chi$ considering attenuation effect numerically for different interactions like constant (magenta), scalar (red solid) and vector mediated (blue dashed dot) cross-sections depicted by different colors. The unattenuated flux (grey) and analytically obtained attenuated flux for constant cross-sections (black dotted) are also shown for comparison. All the results are shown for $m_\chi=1~\mathrm{MeV}$ and $\bar{\sigma}_{\chi e}=10^{-31}~\mathrm{cm}^2$.
    As source we consider PBH with $M_{\rm PBH} =10^{15}~{\rm gm}$ and $ f_{\rm PBH}=3.9\times 10^{-7}$.}
    \label{fig:flux_att}
\end{figure}

{\bf Step 2.} Following the same discussion we now proceed with the second step to obtain the flux. We display the variation of the differential flux of $\chi$ with $T_\chi^z$ in Fig.~\ref{fig:flux_att}.
We consider a PBH with $M_{\rm PBH} =10^{15}~{\rm gm}, f_{\rm PBH}=3.9\times 10^{-7}$ and a benchmark of DM parameters $m_\chi=1~\mathrm{MeV}$ and $\bar{\sigma}_{\chi e}=10^{-31}~\mathrm{cm}^2$.
We take $h_d=1.4$ km, which corresponds to the depth of XENONnT.
For a constant cross-section, we show the fluxes obtained both numerically (magenta solid) and analytically (black dashed) along with the fluxes for scalar (red solid) and vector mediated (blue dashed dot) cross-sections after including attenuation effect (numerically). For comparison we also portray the unattenuated flux. 
Note that even with typical cross-section values $\bar{\sigma}_{\chi e}\ll 10^{-29}{~\rm cm}^2$,
the effect of attenuation is prominent. 
For the constant cross-section we find an analytical treatment (black dashed) of the attenuation leads to an underestimation of the flux at higher energies compared to the actual numerical one (magenta solid). On the other hand, the attenuated fluxes for scalar (red solid) and vector (blue dashed dot) cross-sections also differ significantly from the constant one as well as from the unattenuated flux. It is worth highlighting that for all interaction types, attenuation decreases the number of $\chi$ flux in the higher energy range and consequently the flux increases at lower KE. This feature plays a key role in constraining the DM parameter space, as will be discussed shortly.
The minimum KE of $\chi$ required to generate recoil $E_R\sim 10$ keV is indicated by the orange solid line in the same plot.

\begin{figure}[!tbh]
    \centering
    \includegraphics[scale=0.5]{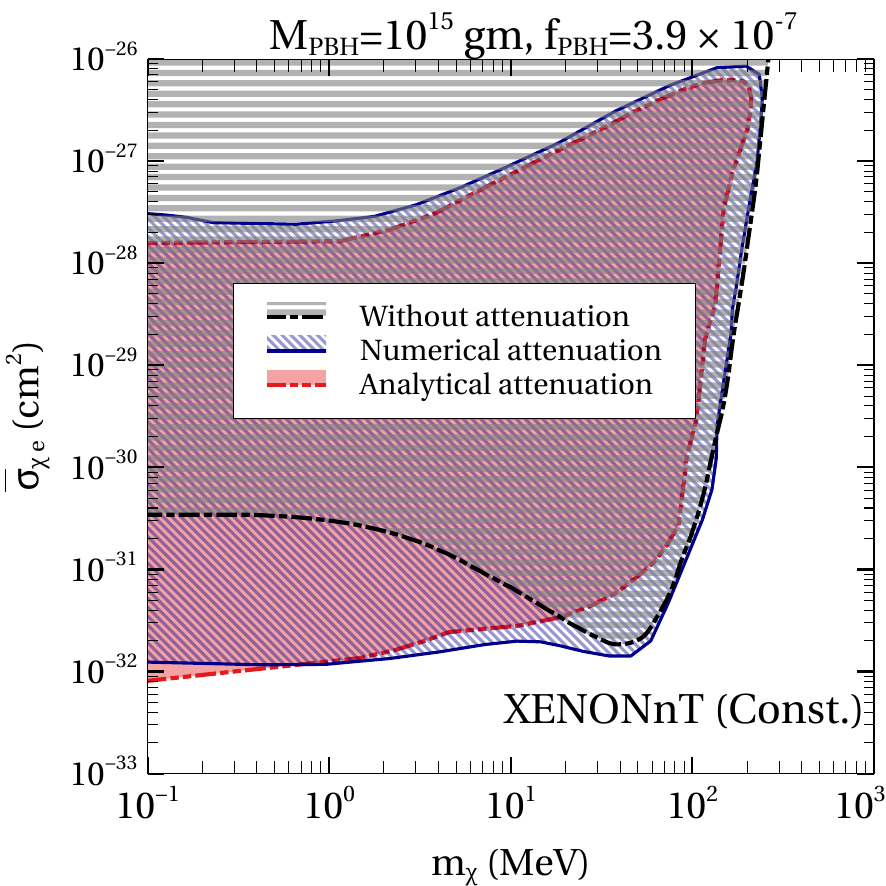}
    \caption{{\bf Step 3.} Comparison between the $2\sigma$ C.L. constraints obtained for constant interaction in three different scenarios-(i) without considering attenuation, (ii) after considering attenuation analytically and (iii) by numerically solving the energy loss equation. Note that consideration of attenuation leads to almost $\mathcal{O}(1)$ stronger limit at $m_\chi\lesssim 1$ MeV. For $m_\chi\gtrsim 20$ MeV, the limit obtained with numerical treatment of attenuation is stronger by factor $\sim 3$ than that  with  analytical treatment.}
    \label{fig:const_compare}
\end{figure}

{\bf Step 3.} We now approach the third step prescribed earlier to give a more rigorous realization.
First, we find the event spectra for fixed $m_\chi$ and $\bar{\sigma}_{\chi e}$  following the usual methodology prescribed in Sec.~\ref{Sec:Event_Simulation}, but using the attenuated DM flux evaluated in the earlier step.
A detailed discussion of the event spectra after considering attenuation effect can be found in Appendix \ref{apx:A}.
The we do a full parameter space scan in the $m_\chi$ vs. $\bar{\sigma}_{\chi e}$ for a constant cross-section (given in eq.~\eqref{eq:sig_const}) and show the constraints from XENONnT in Fig.~\ref{fig:const_compare}.
To place the constraints we follow the same satistical analysis presented in Sec.~\ref{Sec:Event_Simulation}.
The different colored regions in the aforementioned figure represent the constraints at $2\sigma$ C.L. on $\chi$ evaporated from a PBH with $M_{\rm PBH} =10^{15}~{\rm gm}, f_{\rm PBH}=3.9\times 10^{-7}$ from the null observation in XENONnT. 
The region shown with horizontal gray hatched region represents the parameter space excluded when we do not include the effect of attenuation. 
On the other hand, the red (blue backward hatched) region signifies the excluded parameter space after considering analytical (numerical) attenuation.
The attenuation effect places an upper ceiling in the constraints on $\bar{\sigma}_{\chi e}$ as expected.
For example, for $m_\chi=1$ MeV with $\bar{\sigma}_{\chi e}\gtrsim\mathcal{O}(10^{-28}){\rm~cm}^2$ the mean free path of $\chi$ becomes $\ll \mathcal{O}(1)$ km, the typical detector depth and physically the $\chi$ particles scatter significantly before reaching the detector loosing its KE and leading to a reduced flux of $\chi$ at the detector.
Thus the underground detectors technically fails to constrain $\chi$ with higher interaction strength i.e.  $\bar{\sigma}_{\chi e}\gtrsim\mathcal{O}(10^{-28}){\rm~cm}^2$ shown by the upper boundaries of the red and blue regions.
This is a typical feature in direct searches of boosted dark sector particles; thus, considering the attenuation effect is important for studying such scenarios, including PBH-evaporated boosted dark sectors.

Apart from that, the other most crucial aspect of considering attenuation effect can be observed from the lower $\bar{\sigma}_{\chi e}$ region (i.e. in the upper limits $\chi$ interaction) in Fig.~\ref{fig:const_compare}.
Consideration of attenuation effect leads to  almost one order stronger constraint in the upper limit of $\bar{\sigma}_{\chi e}$ especially at low mass region ($m_\chi \lesssim \mathcal{O}(10)$ MeV).
The reason can be easily apprehended from the discussion in the context of Fig.~\ref{fig:flux_att}. Attenuation decreases the number of $\chi$ with higher KE and hence increases the flux at lower KE. 
Thus if $\chi$ has KE $> T_\chi^{\min}(E_R\gtrsim E_{\rm th})$, consideration of attenuation effect leads to an increase in the recoil event in the low energy range compared to the case without attenuation.
The effect is more prominent in the low mass range.
Thus we highlight that attenuation enhances the recoil rate and hence the constraint for the boosted light particles (with constant interactions) emitted from PBH, justifying the importance of considering attenuation.

After successfully demonstrating the attenuation effect we now compare the results obtained after considering attenuation (i) analytically (red) and (ii) numerically (blue) in the context of Fig.~\ref{fig:const_compare}.
Our observations from the aforementioned figure are as follows:
\begin{enumerate}
    \item We find that treating attenuation effect analytically (red) leads to weaker constraints on the upper limit of $\bar{\sigma}_{\chi e}$ for $m_\chi \gtrsim \mathcal{O}(1)$ MeV than the numerical treatment (blue). This can also be explained from Fig.~\ref{fig:flux_att} where we see analytical treatment leads to under estimation of flux at higher KE compared to numerical treatment. 
    \item In the low mass regime $m_\chi \lesssim \mathcal{O}(1)$ MeV analytical treatment leads to overestimation of recoil events, leading to erroneous stronger upper limit on $\bar{\sigma}_{\chi e}$ than numerical treatment.
    \item Also in upper ceiling in the constraint region, numerical treatment leads to almost $\sim 2$ factor stronger constraint on $\bar{\sigma}_{\chi e}$ for $m_\chi \lesssim \mathcal{O}(1)$ MeV.
\end{enumerate}
Following the discussions made above, even though analytical treatment of attenuation effect is possible for constant interaction of $\chi$ we prefer the numerical treatment for more accurate limits.
For scalar and vector-mediated interactions we are not aware of any analytical treatment which leaves us with no option but to solve it numerically.
For all these reasons, throughout the rest part of this paper we consider the effect of attenuation numerically (unless mentioned explicitly) and place the constraints on $\chi$ evaporated from PBH which has not been yet covered in the existing literature to the best of our knowledge.

\section{Results with attenuation effect}
\label{sec:results}
\begin{figure}[!tbh]
    \centering
    \subfigure[\label{ca1}]{
    \includegraphics[scale=0.5]{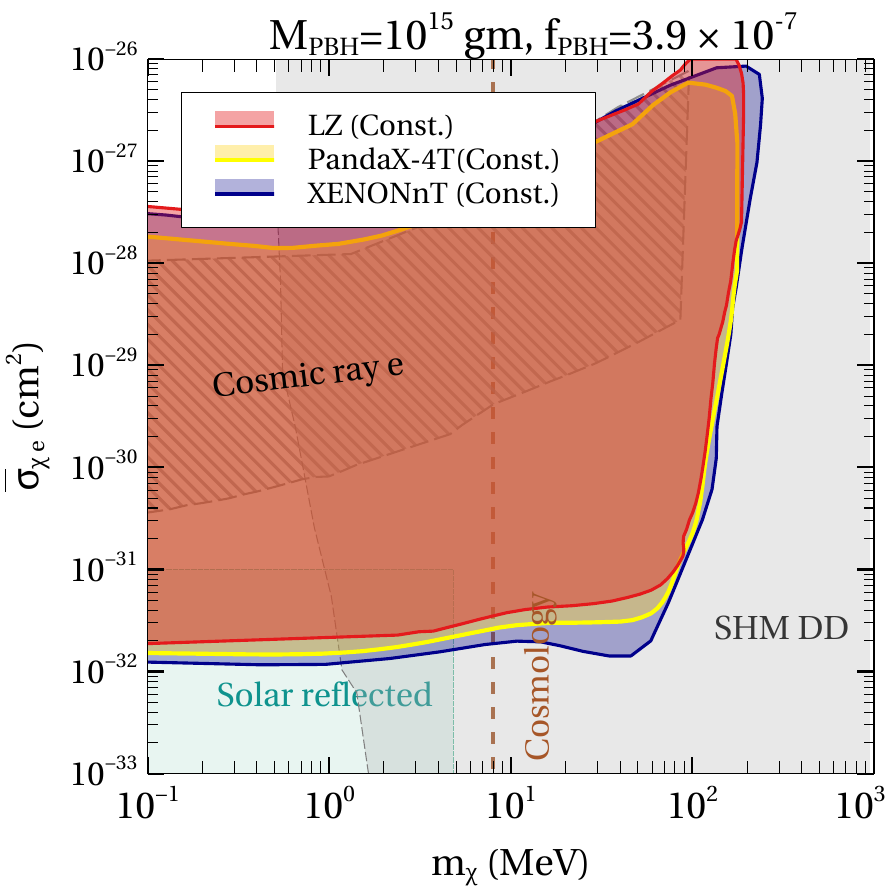}}
    \subfigure[\label{ca2}]{
    \includegraphics[scale=0.5]{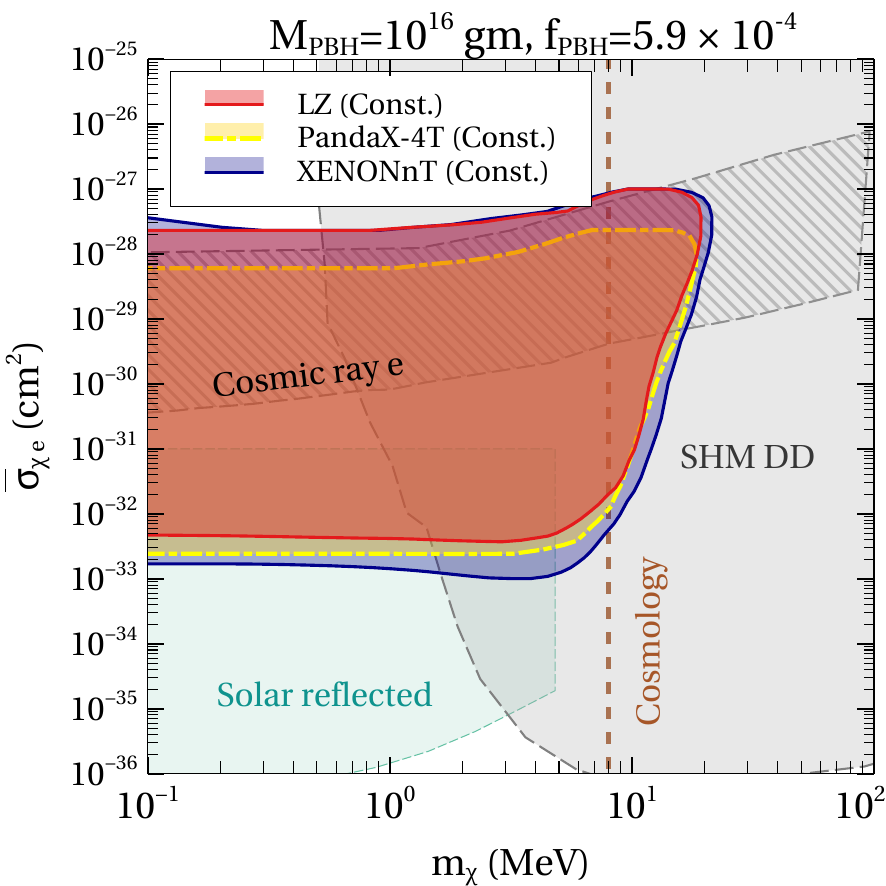}}
    \caption{{\bf Constant cross-section:} Constraints at $2\sigma$ C.L. on DM $\chi$ with constant cross-section evaporated from PBH after {\bf including attenuation effect}, from XENONnT (light blue), PandaX-4T (light yellow) and LZ (light red) . Chosen PBH BPs are (a) $M_{\rm PBH}=10^{15}$ gm with $f_{\rm PBH}=3.9\times 10^{-7}$ and (b) $M_{\rm PBH}=10^{16}$ gm with $f_{\rm PBH}=5.9\times 10^{-4}$. Other existing constraints are also displayed. }
    \label{fig:con_att}
\end{figure}

In the previous section, we argue that attenuation is one of the most crucial effects while studying any boosted dark sector recoil.
In this section, we present the constraints on PBH evaporated DM upon experiencing Earth's attenuation effect \footnote{Detailed discussion of the event spectra after considering attenuation effect can be found in Appendix \ref{apx:A}.}.
We display the constraints from DM detectors only e.g. XENONnT, LZ and PandaX-4T shown by light blue, light red and light yellow regions and 
throughout this section we follow this color combination. Note that, all the constraints presented in this section are obtained, including  numerical attenuation.
For the source PBH of $\chi$ again we consider the earlier mentioned two BPs: (1)$M_{\rm PBH}=10^{15}$gm with $f_{\rm PBH}=3.9\times 10^{-7}$ and $M_{\rm PBH}=10^{16}$ gm with $f_{\rm PBH}=5.9\times 10^{-4}$ allowed from the current constraints on PBH \cite{Malyshev:2023oox}.
First, we discuss the contours assuming constant interaction between $\chi$ and $e$ shown in Fig.~\ref{fig:con_att} considering $M_{\rm PBH}=10^{15}$ gm (Fig.~\ref{ca1}) and $M_{\rm PBH}=10^{16}$ gm (Fig.~\ref{ca2}). For comparison, we also portray other existing constraints on boosted light DM e.g. from cosmic electron boosted DM in XENON1T \cite{Bardhan:2022bdg} (grey hatched region) and solar reflected DM \cite{An:2017ojc} (light cyan region).
The combined limit for galactic halo DM assuming standard halo model (SHM) from SuperCDMS \cite{SuperCDMS:2018mne}, SENSEI \cite{Crisler:2018gci}, and XENON10 \cite{Essig:2012yx} is shown by the grey shaded region.
However, if  $\chi$ is a generic long-lived particle, then these DM constraints are not applicable in our analysis as mentioned earlier. We also include the cosmological bound on fermion DM from BBN \cite{Krnjaic:2019dzc} shown by brown dashed line.

\begin{figure}[!tbh]
    \centering
    \subfigure[]{
    \includegraphics[scale=0.5]{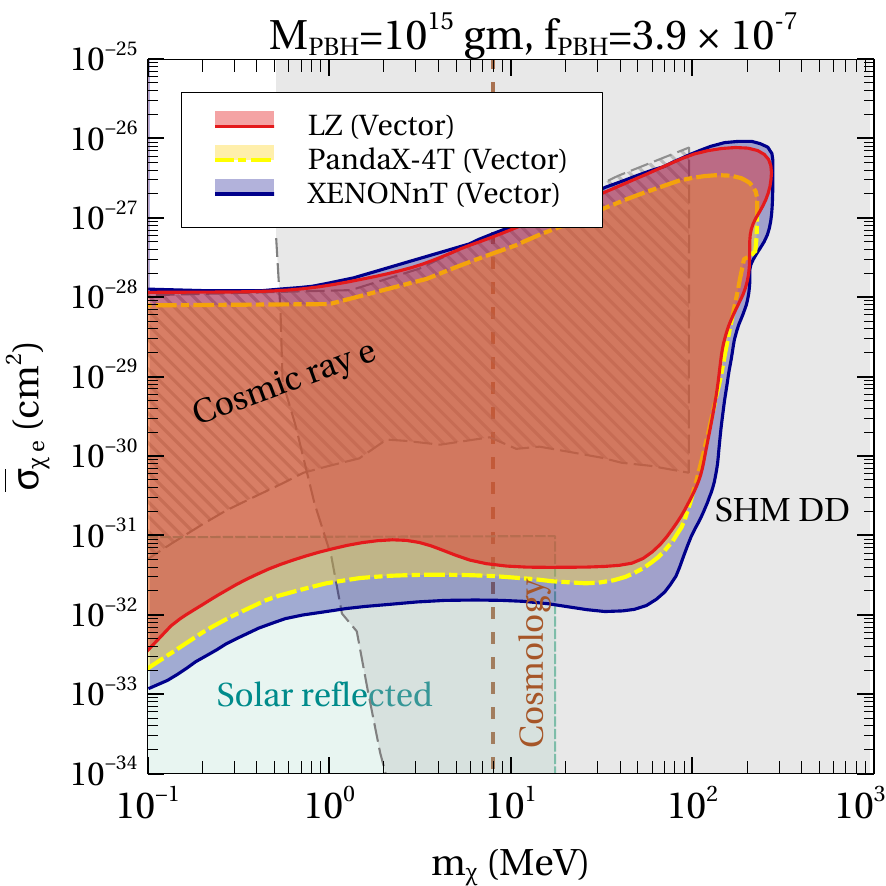}}
    \subfigure[]{
    \includegraphics[scale=0.5]{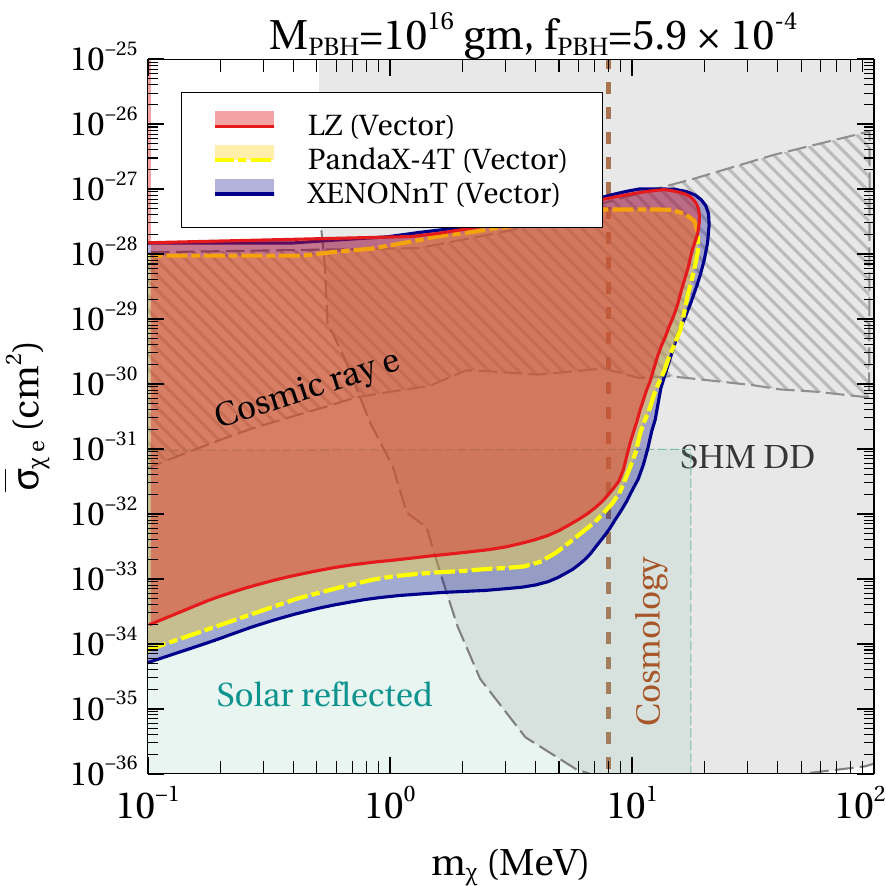}}
    \caption{{\bf Vector cross-section:} Constraints  at $2\sigma$ C.L. on DM $\chi$ with vector cross-section evaporated from PBH after {\bf including attenuation effect}, from XENONnT (light blue), PandaX-4T (light yellow) and LZ (light red) . Chosen PBH BPs are (a) $M_{\rm PBH}=10^{15}$ gm with $f_{\rm PBH}=3.9\times 10^{-7}$ and (b) $M_{\rm PBH}=10^{16}$ gm with $f_{\rm PBH}=5.9\times 10^{-4}$. Other existing constraints are also displayed. }
    \label{fig:vec_att}
\end{figure}
\begin{figure}[!tbh]
    \centering
    \subfigure[]{
    \includegraphics[scale=0.5]{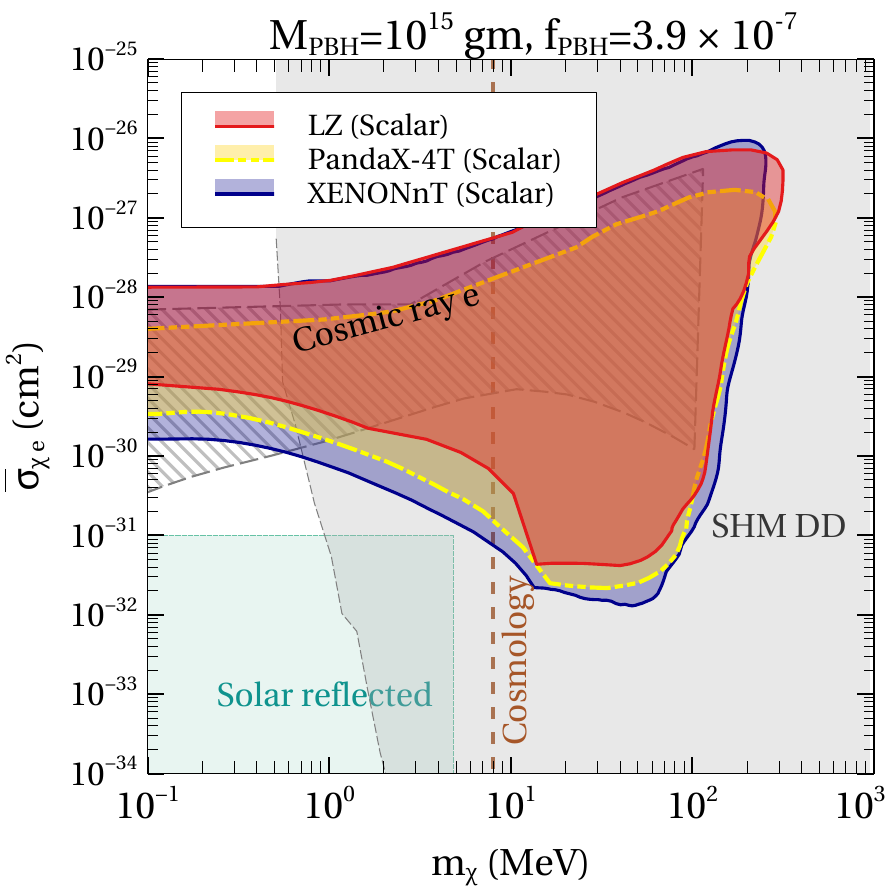}}
    \subfigure[]{
    \includegraphics[scale=0.5]{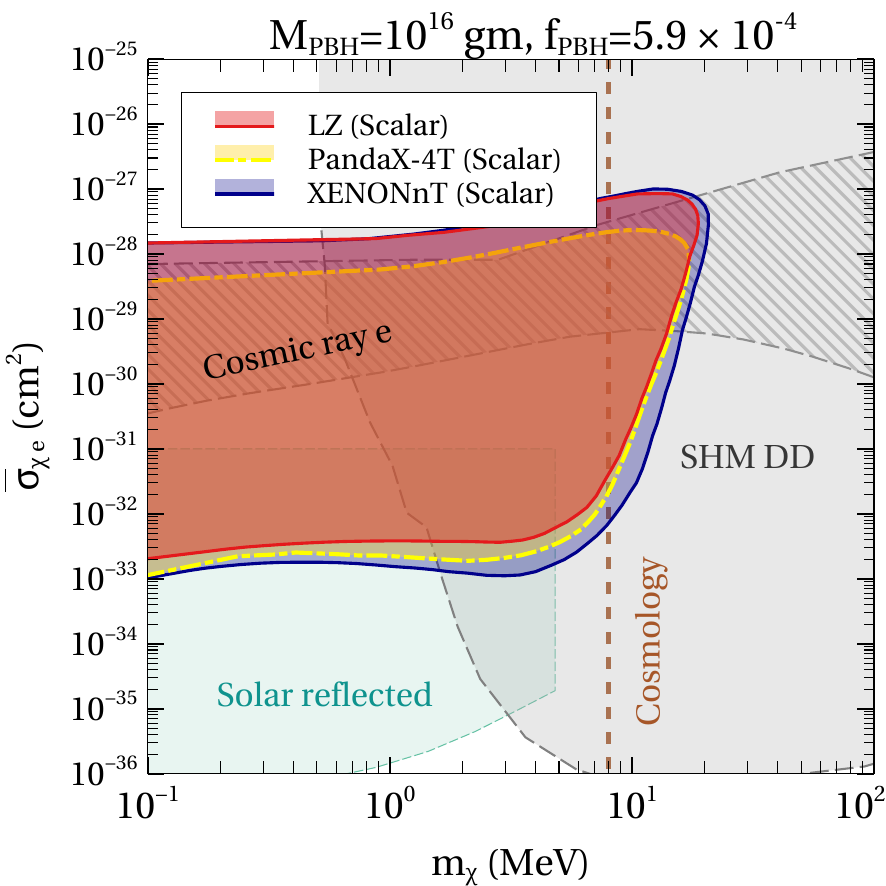}}
    \caption{{\bf Scalar cross-section:} Constraints at $2\sigma$ C.L. on DM $\chi$ with scalar cross-section evaporated from PBH after {\bf including attenuation effect}, from XENONnT (light blue), PandaX-4T (light yellow) and LZ (light red) . Chosen PBH BPs are (a) $M_{\rm PBH}=10^{15}$ gm with $f_{\rm PBH}=3.9\times 10^{-7}$ and (b) $M_{\rm PBH}=10^{16}$ gm with $f_{\rm PBH}=5.9\times 10^{-4}$. Other existing constraints are also displayed. }
    \label{fig:scal_att}
\end{figure}

For both the chosen BPs XENONnT places the strongest constraint, followed by PandaX-4T and LZ, respectively. This hierarchy is due to the larger exposure for XENONnT. If $\chi$ is assumed to be a DM, these limits become competitive to existing other limits and even stronger in the region $m_\chi\lesssim 1$ MeV and $10^{-31}{\rm cm}^2 \lesssim \bar{\sigma}_{\chi e}\lesssim 4\times 10^{-31}{\rm cm}^2$. As hinted earlier due to the attenuation effect, we see upper ceilings in all our obtained limits.
Note that since PandaX-4T has higher over burden ($h_d=2.4$ km) compared to XENONnT and LZ, we observe the ceiling for PandaX-4T is weaker by almost a factor 2.

Similarly, we show the obtained limits for vector and scalar interactions in Fig.~\ref{fig:vec_att} and Fig.~\ref{fig:scal_att} respectively for the chosen PBH BPs $M_{\rm PBH}=10^{15}$ gm ({\bf left pannels}) and $M_{\rm PBH}=10^{16}$ gm ({\bf right pannels}). To denote the constraints we follow the same color combination as used in the constant case. However, the existing constraints on light DM e.g cosmic ray boosted DM \cite{Bardhan:2022bdg} differ for different interactions. The solar reflected DM bound for vector interaction is taken from Ref.~\cite{Emken:2024nox}.
Since we do not find the solar reflected DM bound for scalar interaction in existing literature we portray the existing bound for constant cross-section \cite{An:2017ojc} as shown in Fig.~\ref{fig:scal_att}. 

\section{Constraints on PBH fraction}
\label{sec:fpbh}

\begin{figure}[!tbh]
    \centering
    \subfigure[]{
    \includegraphics[scale=0.5]{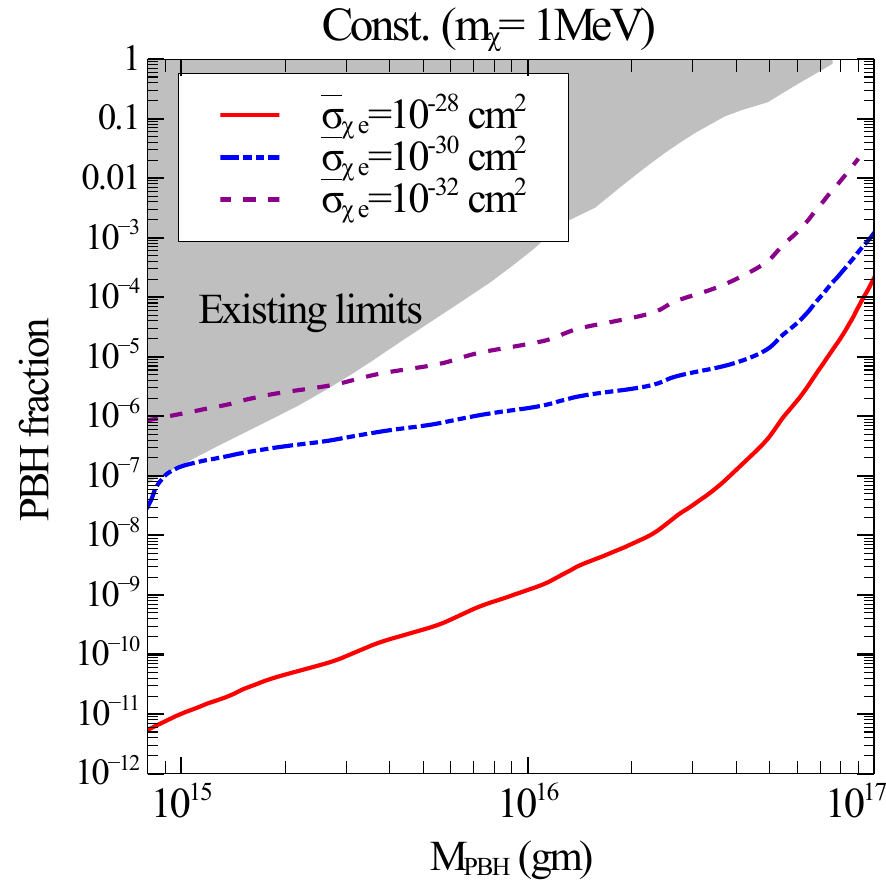}}
    \subfigure[]{
    \includegraphics[scale=0.5]{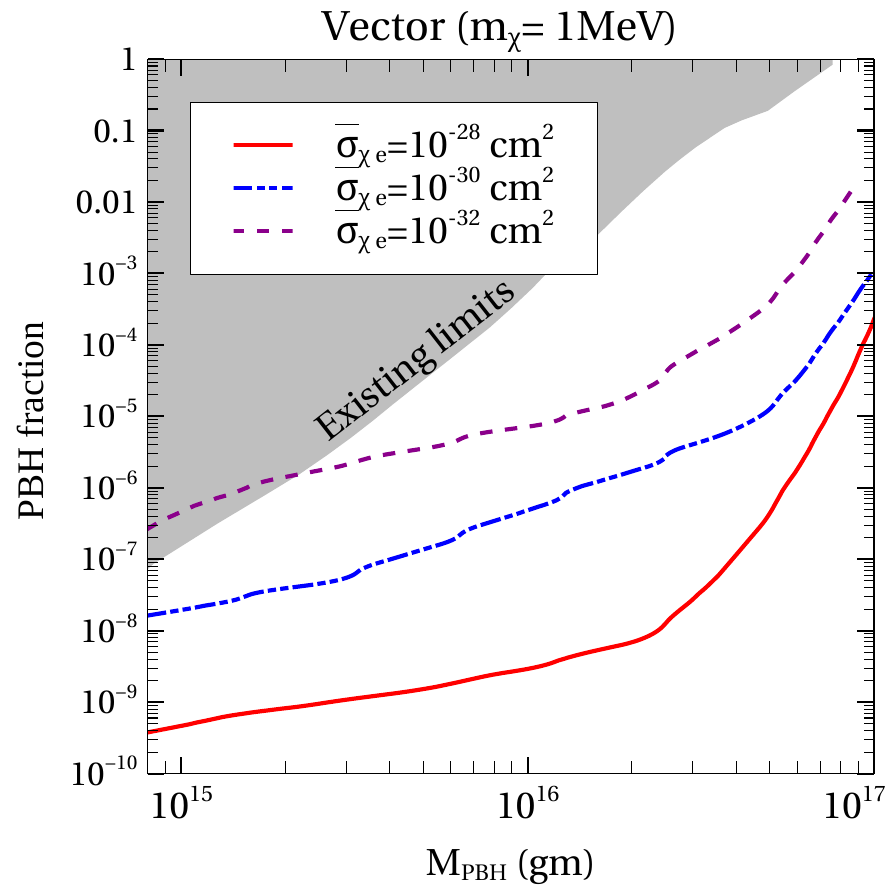}}
    \caption{Constraints  at $2\sigma$ C.L. in $M_{\rm PBH}$ vs. PBH fraction, $f_{\rm PBH}$, from the null observation of a $1$ MeV DM $\chi$  evaporated from PBH after {\bf including attenuation effect}, from XENONnT with (a) constant and (b) vector cross-section. Other existing constraints~\cite{Carr:2020gox} are also displayed.}
    \label{fig:f_pbh}
\end{figure}

The constraints showcased in the previous section, where limits were presented in the DM parameter space spanned by the DM--electron scattering cross-section and the DM mass for benchmark choices of $f_{\rm PBH}$ and $M_{\rm PBH}$, can be straightforwardly reinterpreted as upper bounds on the PBH abundance. In particular, these results can be translated into constraints on the PBH fraction $f_{\rm PBH}$ as a function of the PBH mass $M_{\rm PBH}$. For example, we show the re-casted limits in the $M_{\rm PBH}$ vs. $f_{\rm PBH}$ plane in Fig.~\ref{fig:f_pbh} for a benchmark DM mass of $m_\chi = 1~\mathrm{MeV}$ and a different values of $\bar{\sigma}_{\chi e}$, considering both a constant DM--electron scattering cross-section (left panel) and a vector-mediated interaction (right panel).
These limits are obtained
using XENONnT data and including the effects of DM flux attenuation. While similar limits can also be obtained using LZ and PandaX-4T data, the results of the previous section indicate that XENONnT currently provides the strongest sensitivity; for this reason, we restrict the discussion here to XENONnT-derived constraints. For comparison and complementarity, we also overlay existing bounds in the $M_{\rm PBH}$--$f_{\rm PBH}$ plane, combining constraints from PBH evaporation such as those from Big Bang nucleosynthesis~\cite{Carr:2009jm}, cosmic microwave background spectral distortions and anisotropies~\cite{Acharya:2020jbv, Chluba:2020oip}, intergalactic medium (IGM) thermal history~\cite{Khan:2025kag}, global 21-cm signal measurements~\cite{Mittal:2021egv}, extragalactic~\cite{Carr:2009jm} and Galactic $\gamma$-ray observations~\cite{Carr:2016hva}, and Voyager-1 $e^\pm$ measurements~\cite{Boudaud:2018hqb}. From Fig.~\ref{fig:f_pbh}, it is evident that XENONnT yields competitive, and in certain regions superior, constraints on $f_{\rm PBH}$ even for comparatively small DM--electron cross-sections, down to $\bar{\sigma}_{\chi e}\sim10^{-32}~\mathrm{cm}^2$, which remains allowed by current XENONnT constraints, particularly for $M_{\rm PBH} \gtrsim 3\times10^{15}~\mathrm{gm}$. Finally, we emphasize that the bounds on $f_{\rm PBH}$ and $\bar{\sigma}_{\chi e}$ are obtained simultaneously within the same analysis and are therefore intrinsically correlated.

\section{Prospects in neutrino detectors}
\label{sec:nu_detector}

\begin{figure}[!tbh]
    \centering
    \subfigure[\label{s1}]{\includegraphics[scale=0.5]{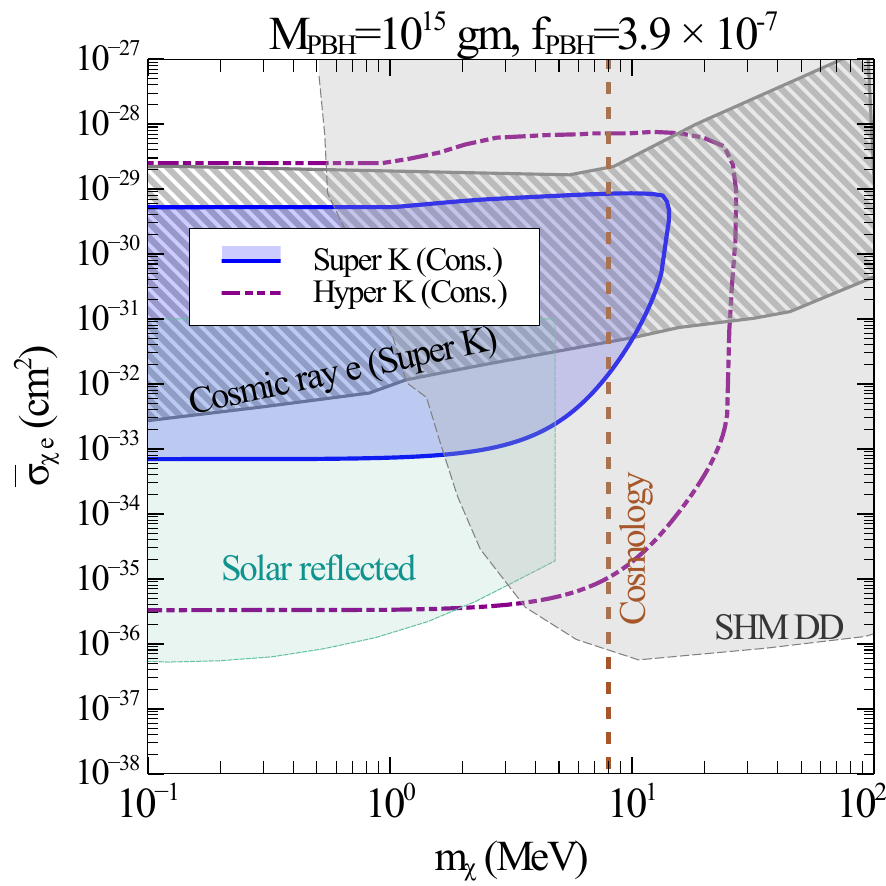}}
    \subfigure[\label{s2}]{\includegraphics[scale=0.5]{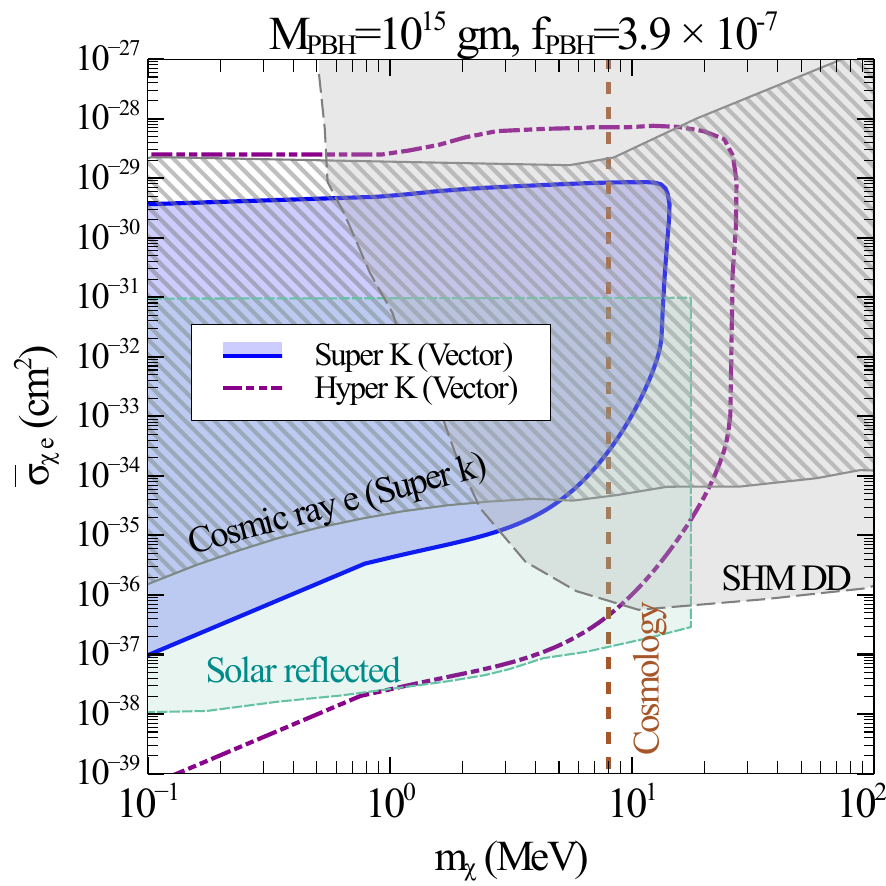}}
    \subfigure[\label{s3}]{\includegraphics[scale=0.5]{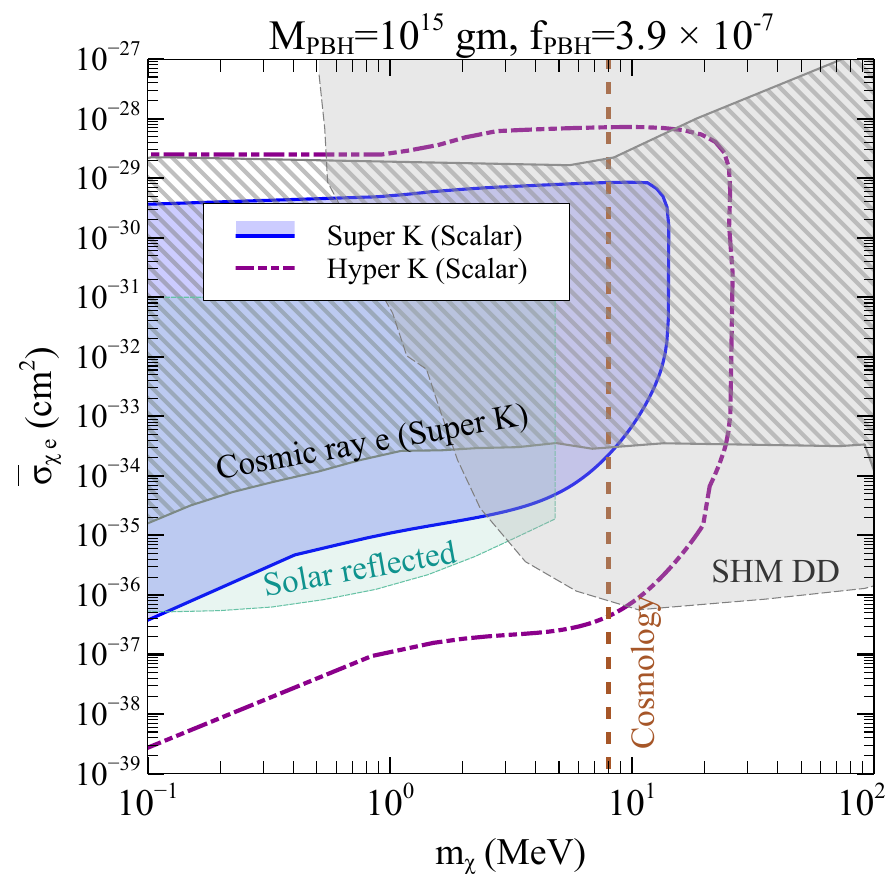}}
    \caption{Conservative constraints on DM $\chi$ evaporated from PBH with $M_{\rm PBH}=10^{15}$ gm with $f_{\rm PBH}=3.9\times 10^{-7}$ from Super-K (light blue) and Hyper-K (dashed dot dot magenta line) for (a) constant, (b) vector and (c) scalar cross-sections of $\chi$. Other existing constraints are also displayed. }
    \label{fig:Sup-k}
\end{figure}

Having a detailed analysis with DM detectors, we now briefly discuss the constraints and projected reach from existing and future generation multi-tonne-scale neutrino detector like Super-Kamiokande (Super-K) and Hyper-Kamiokande (Hyper-K). Super-K is a 50 kilotonne water Cherenkov detector \cite{Super-Kamiokande:2017dch}. The large target electron density ($N_e=3.34\times 10^{28}~{\rm tonne}^{-1}$) in Super-K helps to place stronger constraints on light DMs compared to DM detectors \cite{Super-Kamiokande:2017dch,Ema:2018bih}.
Super-K data in \cite{Super-Kamiokande:2017dch} corresponds to 2628.1 days of data taking with 161.9 kilotonne-year exposure.
They reported 4042 events observed in the first bin with $0.1~{\rm GeV}< E_R<1.33$~GeV with an efficiency $\mathcal{\epsilon}_{\rm SK}=0.93$.
To obtain the Super-K limits we follow the method used in Ref.~\cite{Ema:2018bih,Calabrese:2022rfa} where we consider the predicted event in the aforementioned energy range $N_{\rm SK}>4042$ to place a conservative limit i.e.
\begin{eqnarray}
   n_T~ t_{\rm run} ~\mathcal{\epsilon}_{\rm SK}\times \int_{0.1{~\rm GeV}}^{1.33{\rm ~GeV}} \frac{\diff R}{\diff E_R} \diff E_R > 4042,
\end{eqnarray}
with $n_T~ t_{\rm run}=7.5\times 10^{34}.$year \cite{Ema:2018bih}.
Note that we do not consider the bins in the higher energies as the maximum KE $\chi$ can have is $\sim 300$ MeV (see Fig.~\ref{fig:uan_flux}) for our chosen PBH BPs.
It is also crucial to note that with this Super-K data \cite{Super-Kamiokande:2017dch} one can not constrain $\chi$ evaporated from a PBH with mass $m_{\rm PBH}=10^{16}$~gm since the $T_\chi$ always falls below the chosen threshold 0.1 MeV ((see Fig.~\ref{fig:uan_flux})).

We show our obtained constraints from Super-K on $\chi$ in Fig.~\ref{fig:Sup-k} considering 
$M_{\rm PBH}=10^{15}$~gm and $f_{\rm PBH}=3.9\times 10^{-7}$.
Our obtained limits for constant (Fig.~\ref{s1}), vector (Fig.~\ref{s2}) and scalar (Fig.~\ref{s3}) cross-section are depicted with light blue shaded region.
We incorporate the full numerical implementation to include the attenuation effect, following the methodology prescribed in Sec.~\ref{Sec:Attenuation_Technique}.
The attenuation effect in the lower part of the exclusion contour is found to be less severe as Super-K explores smaller ($\lesssim 10^{-33}{~\rm cm}^2$) cross-sections than DM detectors.
However, attenuation imposes a strong ceiling on the cross-section around  $\bar{\sigma}_{\chi e}$ above $\gtrsim 4 \times 10^{-30}{~\rm cm}^2$ as the PBH evaporated DM with such a high cross-section gets significantly attenuated before reaching Super-K detector situated at a depth of $h_d\sim$ 1 km.
Note that the attenuation ceiling for Super-K lies at a smaller cross-section than that corresponding to DM detectors because of 
the higher threshold energy ($\sim 100$ MeV) of Super-K detector compared to DM detectors.
In the same plane, we also showcase the previously mentioned existing limits from
cosmic electron boosted DM in Super-K \cite{Bardhan:2022bdg} (grey hatched region) and solar reflected DM \cite{An:2017ojc} (light cyan region) and
the combined limit for SHM \cite{SuperCDMS:2018mne,  Crisler:2018gci, Essig:2012yx}.

In the same planes in Fig.~\ref{fig:Sup-k} we display the projected reach of Hyper-K shown by magenta dashed dot dot line for different interaction types. 
Hyper-K is a future generation water cherenkov detector expected to have 187 kilotonne fiducial mass~\cite{Hyper-Kamiokande:2018ofw}.
For our analysis, we use the 20 year background simulated in Ref.~\cite{Bell:2020rkw} in the energy ranges between 20 MeV to 500 MeV with a flat efficiency $0.7$ and scale the background to  1 year data (${\rm B}_{\rm HK}$) which is a feasible run time. The lower range is fixed due to the spallation background and the upper energy range is chosen as the maximum recoil energy bin in which PBH evaporated DM can generate a recoil. We obtain the conservative limit with a simplistic approach by imposing event predicted at Hyper-K , $N_{\rm HK}> {\rm B}_{\rm HK}$  without doing bin by bin analysis.

The higher target size and low threshold of Hyper-K is reflected in the obtained limits in Fig.~\ref{fig:Sup-k}.
Since Hyper-K can probe even smaller cross-sections than Super-K, the attenuation effect for the lower part of the (projected) exclusion region is found to be negligible too. 
On the other hand, the attenuation ceiling ($\bar{\sigma}_{\chi e}$ above $\gtrsim 7 \times 10^{-29}{~\rm cm}^2$) on the (projected) exclusion region of Hyper-K is found to be more relaxed than Super-K due to the low threshold ($\sim 20$ MeV) of Hyper-K.
Thus, one can observe that Hyper-K can potentially probe almost $\mathcal{O}(10^{-2})$ smaller cross-sections of PBH evaporated DM $\chi$ compared to Super-K.
In some region, the future limits go through previously unexcluded DM parameter space and thus Hyper-K can potentially shed light on the dark sector phenomenology.
In the future, it would be interesting to perform a bin by bin analysis with multi-tonne scale detectors like Hyper-K, JUNO, DUNE to place aggressive limits on PBH evaporated dark sectors.


\section{Conclusion}
\label{sec:concl}
Current DM detectors with $\sim$ keV threshold energy are ill suited to probe MeV scale light DM from galactic halo due to the small velocity of galactic halo DM.
However, PBH can emit energetic sub-GeV dark matter/ dark sector radiation with energies well above the detector thresholds ($\sim1$ keV) independent of their coupling strength to other particles. Thus, such boosted DM can be potentially detected with current DM detectors.
In this work we have systematically studied the detection prospects of such a dark particle emitted from PBH in dedicated DM detectors. 
Null observation of any statistically significant event in such detectors will lead to constraint on the mass and interaction cross-section of such particles. Using the updated results from XENONnT, LZ and PandaX-4T we try to probe these dark sector particles.
Alternately, through the (non) detection of such energetic dark particles one can indirectly probe the PBH present in the universe.  

For a boosted dark particle, with energy dependent scattering cross-sections the event rates and consequently the constraints differ significantly compared to that with a constant cross-section \cite{Bardhan:2022bdg}. Since the fundamental interactions of any dark sector particles is yet unknown one should consider all allowed possibilities.
Keeping this in mind we have considered a dark sector fermion $\chi$ with not only constant $\bar{\sigma}_{\chi e}$ but also with scalar and vector cross-sections with $e$. 
We have considered two fixed BPs for the source PBH (1)$M_{\rm PBH}=10^{15}$ gm with $f_{\rm PBH}=3.9\times 10^{-7}$ and (2) $M_{\rm PBH}=10^{16}$ gm  with $f_{\rm PBH}=5.9\times 10^{-4}$. Both BPs are allowed from the existing limits \cite{Malyshev:2023oox}. 
The flux of $\chi$ from a PBH is calculated using the publicly available code \texttt{BlackHawk v2.3} code~\cite{Arbey:2021mbl}.
Considering such PBH evaporated $\chi$, we calculate the event rate in the ground based low energy ($\sim $keV) sensitive DM detectors e.g. XENONnT, LZ, PandaX-4T and then perform $\chi^2$ analysis to place $2\sigma$ C.L. constraints on the $m_\chi$ vs. $\bar{\sigma}_{\chi e}$ plane (see Sec.~\ref{sec:result_unatt}).
The qualitative feature along with order of magnitude estimation of the constraints on $\chi$ emitted from PBHs with other masses and fractions, can be easily deduced from our analysis.

Attenuation is another important aspect when it comes to directly detecting a boosted dark particle with a typically large ($\bar{\sigma}_{\chi e}\gtrsim \mathcal{O}(10^{-32}){~\rm cm}^2$) cross-section. 
The most crucial feature in the detection of boosted sub GeV DM is its high kinetic energy, which the DM might lose while traversing through Earth's  crust on its way to reach the detector. 
Thus neglecting the attenuation may lead to quite different estimation of the direct detection constraints on such boosted DM.
In this work we have carefully considered the attenuation effect for different detectors by solving the energy loss equation of $\chi$ numerically and then integrating over the zenith angle we calculate the true flux of $\chi$ at the detector depth.
We present a dedicated discussion of the attenuation effect in Sec.~\ref{Sec:Attenuation_Technique}.

For a constant cross-section, an analytical treatment of the attenuation is available in the existing literature  \cite{Bringmann:2018cvk} and the same has been considered to constrain PBH evaporated $\chi$ (with constant $\bar{\sigma}_{\chi e}$ only) in previous literature \cite{Li:2022jxo}.
In Fig.~\ref{fig:const_compare} we first compare the bounds obtained from XENONnT, considering the attenuation effect (both analytically and numerically) with the limit obtained without attenuation effect.
We find that besides setting  an expected upper ceiling in the constraint,  the bound with attenuation effect differs significantly in the lower $m_\chi$ regime than the unattenuated one.
For more accuracy, we then compare the bound obtained with the analytic treatment of attenuation with the exact numerically obtained one.
We quantitatively argue that the analytical treatment of the attenuation leads to significant changes in the constraint compared to the numerical one.
For example, we find that, for a constant  $\bar{\sigma}_{\chi e}$
analytical treatment leads to underestimation (overestimation) of the upper limit of $\bar{\sigma}_{\chi e}$ for $m_\chi \gtrsim \mathcal{O}(1)$ MeV ($\lesssim \mathcal{O}(1)$ MeV) than the numerical treatment (Fig.~\ref{fig:const_compare}), justifying the necessity of the numerical treatment of attenuation.
We have explicitly shown our methodology of the numerical treatment of the attenuation effect, which stands out as another novelty of this work.
We highlight that, though we have discussed in the context of a PBH evaporated electrophilic $\chi$, the same methodology can be used for other boosted DM phenomenology \cite{Bringmann:2018cvk,Das:2021lcr}.
Since, for scalar and vector cross-sections, no analytical approximation is available in the existing literature, we adhere to the numerical treatment.

In Sec.~\ref{sec:results} we present the $2\sigma$ C.L. limits on PBH evaporated $\chi$ on the $m_\chi$ vs. $\bar{\sigma}_{\chi e}$ plane from XENONnT, LZ, PandaX-4T after considering full numerical treatment of the attenuation effect.
We highlight that for all three cross-sections i.e. constant (Fig.~\ref{fig:con_att}), vector  (Fig.~\ref{fig:vec_att}) and scalar (Fig.~\ref{fig:scal_att}) the bounds differ significantly from the bounds obtained without attenuation effect (in Sec.~\ref{sec:result_unatt}) even for $\chi$ evaporated from the same PBH.
The effect is more prominent in the low mass regime and thus justifies our argument of the consideration of numerical treatment of attenuation.
If one associates $\chi$ to be the same galactic DM particle, the existing DM constraints are also relevant for our analysis.
Hence, we also portray the halo DM constraints and boosted DM bounds to compare with and find that our obtained limits from DM detectors are competitive with them and even stronger in some parameter space.
However, we stress that since the origin of the $\chi$ considered here is not the galactic halo DM, it can be any generic dark particle not necessarily DM. 
In this case, the DM constraints do not apply here, and our limits stand as the strongest ones for each PBH BP and each $\chi-e$ interaction type.
As a related point, we also discuss how the detection of any boosted light dark particle $\chi$ can help to probe the PBH fraction in the universe (see Sec.~\ref{sec:fpbh}).

Finally, for completeness, we also briefly discuss the detection prospects of such PBH evaporated DM in the current and future neutrino detectors in Sec.~\ref{sec:nu_detector}. 
We consider the possible events induced by $\chi$ in Super-K and Hyper-K.
The neutrino detectors have both pros and cons for the detection of such particles. Due to their larger target volume they can provide the strongest limit on $\bar{\sigma}_{\chi e}$ even superseding the existing boosted DM limits (see Fig.~ \ref{fig:Sup-k}) for the source PBH $M_{\rm PBH}=10^{15}$ gm.
However, these detectors have very high threshold compared to DM detectors and thus for higher masses of source PBH, $\chi$ do not get sufficient KE to be detected in these detectors.
For the source PBH $M_{\rm PBH}=10^{15}$ gm, Super-K provides the most stringent limit on $\bar{\sigma}_{\chi e}$ for lower DM masses ($\mathcal{O}(1)$ MeV) for scalar and vector type cross-section.
We also show the potential reach of Hyper-K, which may allow to probe even smaller cross-sections thanks to its larger target size.

Thus, in this work we perform a detailed analysis of directly detecting a dark sector evaporated form a PBH through the ongoing DM and neutrino detectors.
In future, from the data observed in experiments with larger exposure and higher sensitivities e.g. JUNO, DUNE, Hyper-K, one can probe even smaller cross-sections of light DM and shed light on the previously unexplored DM parameter space.
Also, as mentioned earlier, through the detection of such dark particles, these experiments can also probe the PBH parameters as well.\\

\noindent{\bf Note Added.} Just before the submission of this work, another study exploring similar idea has appeared~\cite{Zhu:2026ush}. However,  the Ref.~\cite{Zhu:2026ush}  considers only energy independent constant cross-section  relying on analytical attenuation. On the other hand, this work deals with realistic energy-dependent cross-sections incorporating full numerical treatment of the Earth's attenuation. 

\section*{Acknowledgment}
SJ would like to thank Rohan Pramanick for computational support.
 The work of SJ is supported
by the National Natural Science Foundation of China (12425506, 12375101, 12090060, and
12090064) and the SJTU Double First Class start-up fund (WF220442604). 
AM acknowledges financial support from the Government of India via the Prime Minister Research Fellowship (PMRF) scheme, (ID: 0401970).
\appendix
\section{Recoil spectra in DM detectors with attenuation effect}
\label{apx:A}
\begin{figure}[!tbh]
    \centering
    \subfigure[\label{xu1}]{
    \includegraphics[scale=0.3]{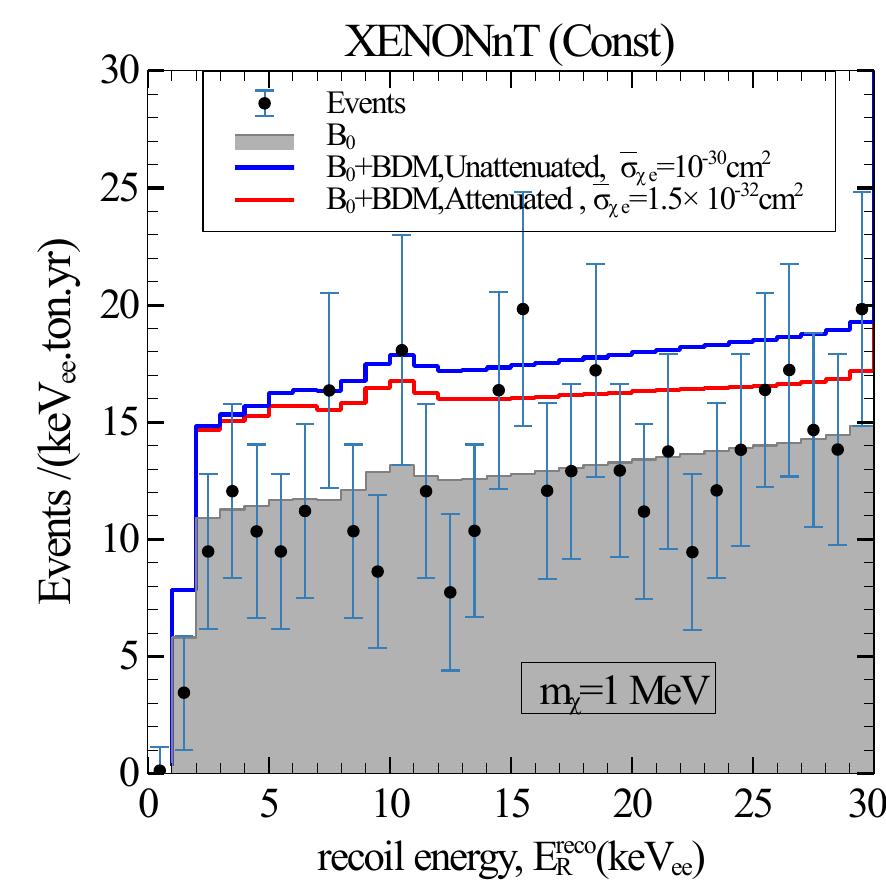}}
    \subfigure[\label{xu2}]{
    \includegraphics[scale=0.3]{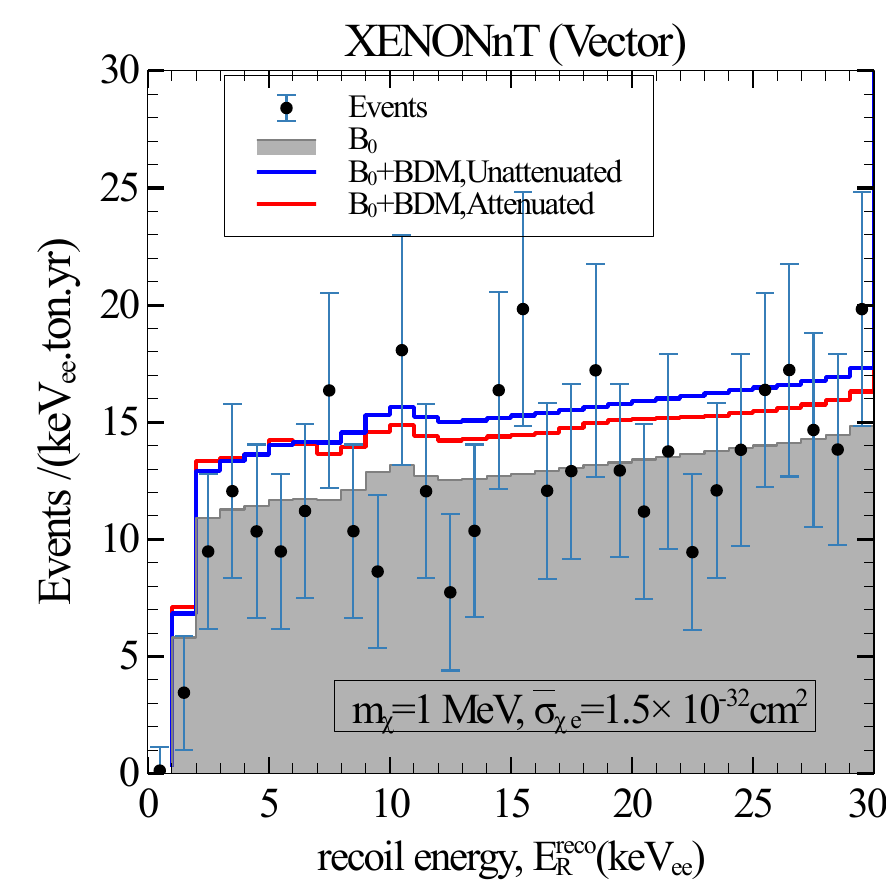}}
    \subfigure[\label{xu3}]{
    \includegraphics[scale=0.3]{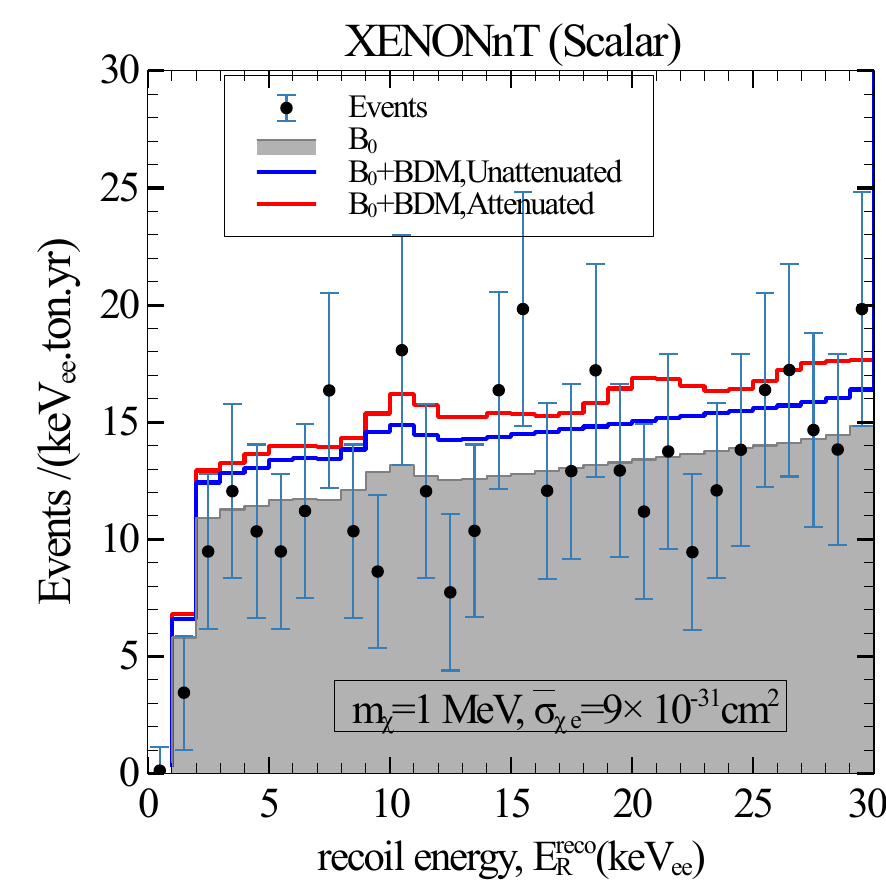}}
    \caption{Expected recoil rate in XENONnT before and after including attenuation effects shown by blue and red lines respectively for $m_\chi=1$ MeV evaporated from a PBH with $M_{\rm PBH}=10^{15}$ gm and $f_{\rm PBH}=3.9 \times 10^{-7}$. Rates for different cross-sections: constant, vector and scalar are shown in (a),(b) and (c) respectively.}
    \label{fig:xe_reco_att}
\end{figure}
\begin{figure}[!tbh]
    \centering
    \subfigure[\label{lu1}]{
    \includegraphics[scale=0.3]{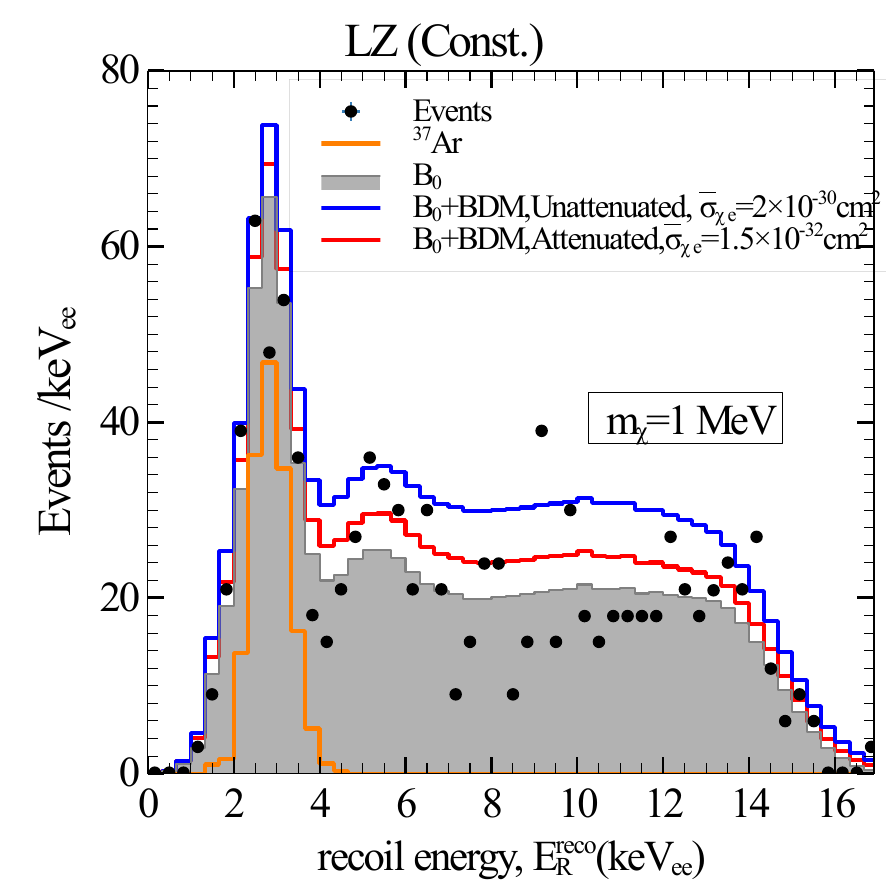}}
    \subfigure[\label{lu2}]{
    \includegraphics[scale=0.3]{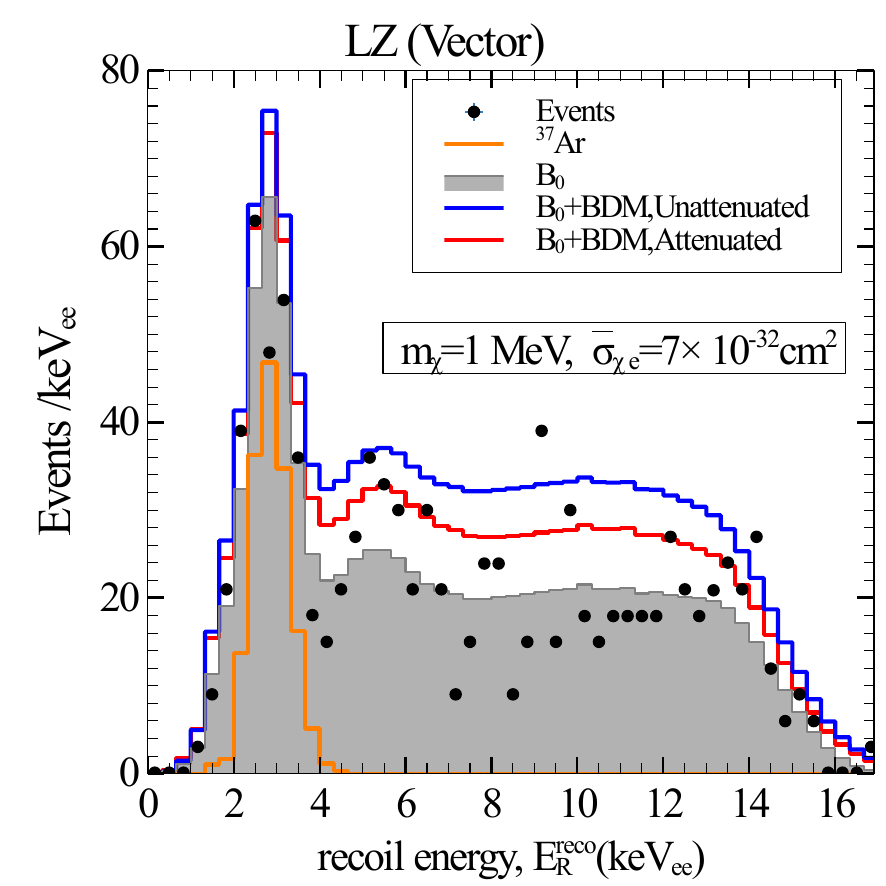}}
    \subfigure[\label{lu3}]{
    \includegraphics[scale=0.3]{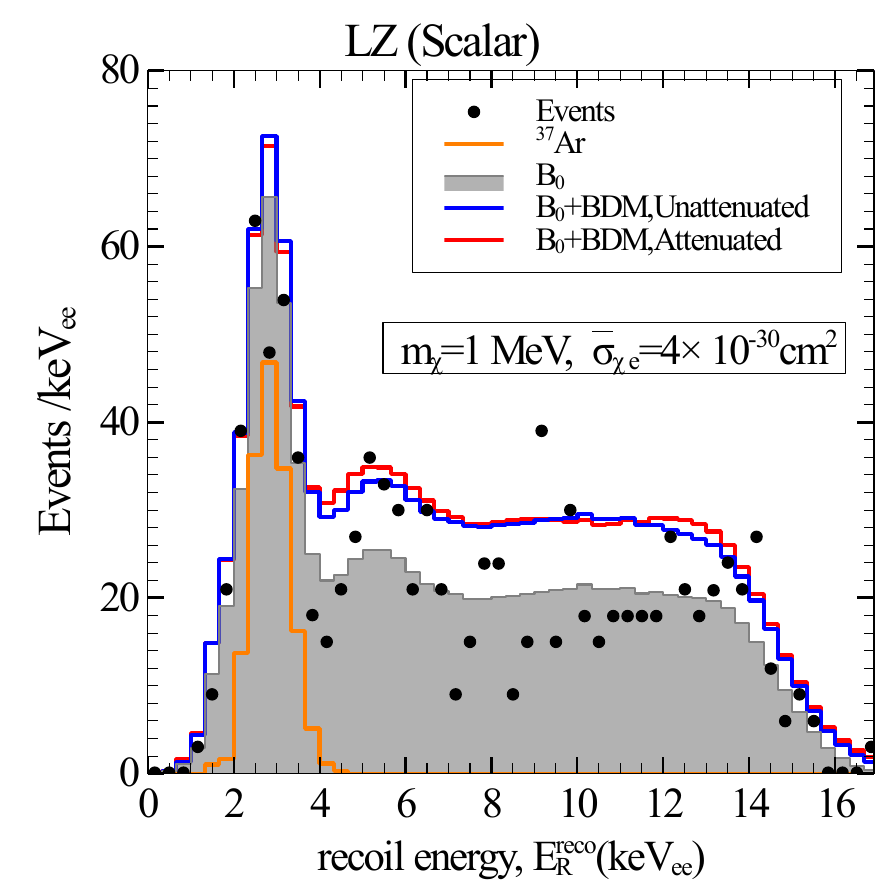}}
    \caption{Expected recoil rate in LZ before and after including attenuation effects shown by blue and red lines respectively for $m_\chi=1$ MeV evaporated from a PBH with $M_{\rm PBH}=10^{15}$ gm and $f_{\rm PBH}=3.9 \times 10^{-7}$. Rates for different cross-sections: constant, vector and scalar are shown in (a),(b) and (c) respectively.}
    \label{fig:lz_reco_att}
\end{figure}
\begin{figure}[!tbh]
    \centering
    \subfigure[\label{pu1}]{
    \includegraphics[scale=0.3]{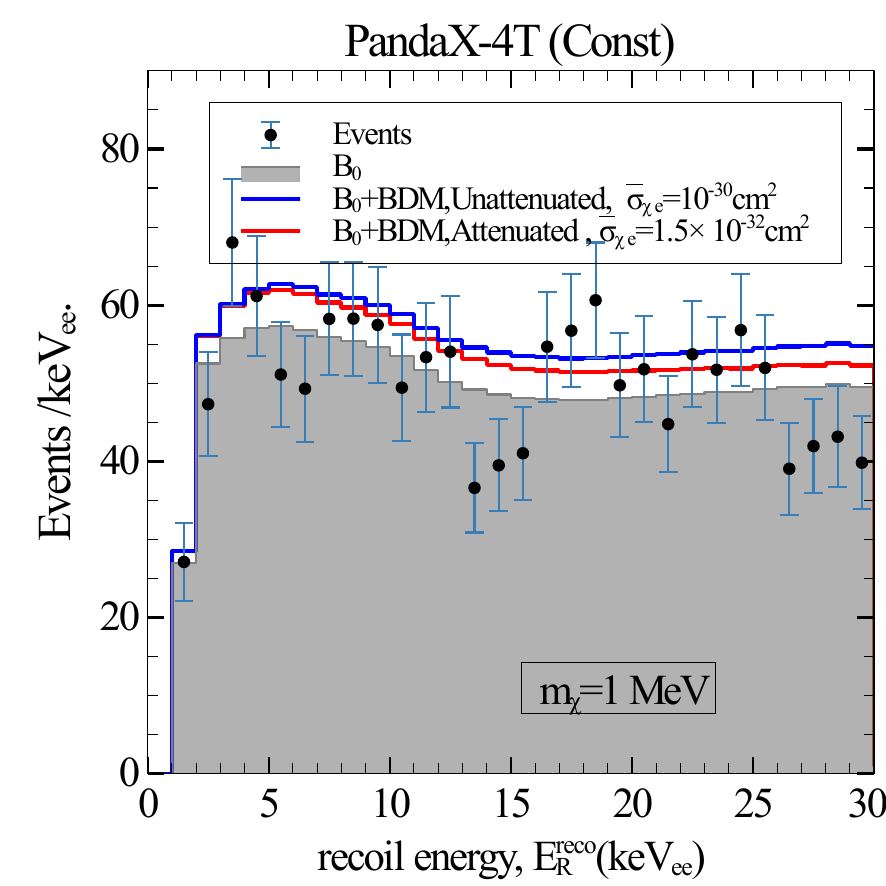}}
    \subfigure[\label{pu2}]{
    \includegraphics[scale=0.3]{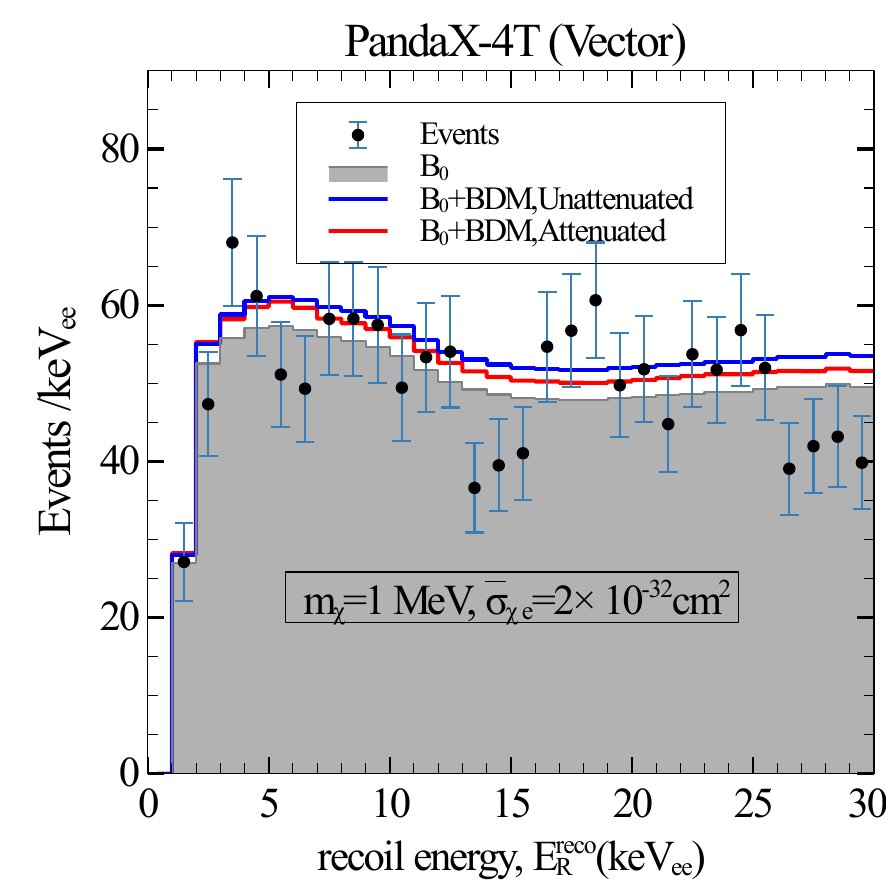}}
    \subfigure[\label{pu3}]{
    \includegraphics[scale=0.3]{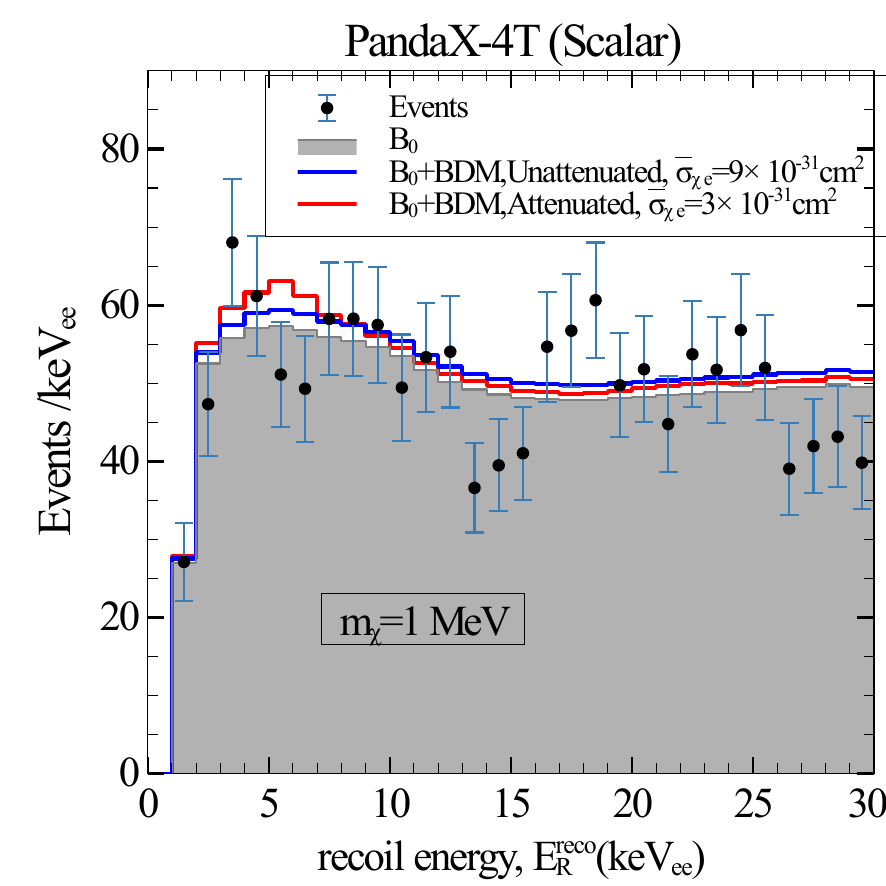}}
    \caption{Expected recoil rate in PandaX-4T before and after including attenuation effects shown by blue and red lines respectively for $m_\chi=1$ MeV evaporated from a PBH with $M_{\rm PBH}=10^{15}$ gm and $f_{\rm PBH}=3.9 \times 10^{-7}$. Rates for different cross-sections: constant, vector and scalar are shown in (a),(b) and (c) respectively.}
    \label{fig:panda_reco_att}
\end{figure}
We showcase the expected recoil signature in XENONnT in Fig.~\ref{fig:xe_reco_att} for $m_\chi =1$ MeV.
The envelope of the grey region signifies the known backgrounds denoted as $B_0$ where as the 
black dots with error bars signify the observed event data.
The red and blue lines signify the recoil rates with and without including attenuation.
Recoil rates for different type of interactions i.e. constant, vector and scalar are shown in Fig.~\ref{xu1}, \ref{xu2} and \ref{xu3} respectively.
For constant cross-section in Fig.~\ref{xu1} we consider $\bar{\sigma}_{\chi e}= 1.5\times 10^{-32}$cm$^2$ (red) and $\bar{\sigma}_{\chi e}= 10^{-30}$cm$^2$ (blue) for cases with and without attenuation, respectively.
For vector (Fig.~\ref{xu2}) and scalar (Fig.~\ref{xu3}) we consider $\bar{\sigma}_{\chi e}= 1.5\times 10^{-32}$cm$^2$ and $\bar{\sigma}_{\chi e}= 9\times 10^{-31}$cm$^2$ respectively (for both the cases: with and without attenuation). Note that after including attenuation, the expected recoil rates change significantly for constant and vector interactions. For constant interaction (Fig.~\ref{xu1}), the change is most prominent where in the attenuated case, a $\mathcal{O}(10^{-2})$ smaller cross-section leads to almost comparable rate to unattenuated case. 

Similarly, we show the expected event rates for LZ in Fig.~\ref{fig:lz_reco_att} for $m_\chi=1$ MeV. 
Here also we follow a similar color convention as in Fig.~\ref{fig:xe_reco_att} with the background due to $^{37}Ar$
shown by orange lines.
For constant cross-section in Fig.~\ref{lu1} we consider $\bar{\sigma}_{\chi e}= 1.5\times 10^{-32}$cm$^2$ (red) and $\bar{\sigma}_{\chi e}= 2 \times 10^{-30}$cm$^2$ (blue) for cases with and without attenuation, respectively.
For vector (Fig.~\ref{lu2}) and scalar (Fig.~\ref{lu3}) we consider $\bar{\sigma}_{\chi e}= 7\times 10^{-32}$cm$^2$ and $\bar{\sigma}_{\chi e}= 4\times 10^{-30}$cm$^2$ respectively (for both the cases: with and without attenuation). 
Our conclusion about the event rates for different interactions remains similar to that in XENONnT, exhibiting the most significant changes for constant interaction after including attenuation.

\FloatBarrier
\bibliography{ref}
\bibliographystyle{JHEP}

\end{document}